\documentclass{article}
\usepackage{mdframed}
\AtBeginDocument{%
  }

\date{}

\newcommand{\countries}[1]{23 {#1}}
\newcommand{\total}[1]{26 {#1}}
\usepackage{tcolorbox}
\newtcolorbox{summarybox}[1][]{
  colback=gray!10,
  colframe=gray!135,
  arc=1mm,
  boxrule=0.5pt,
  left=5pt,
  right=5pt,
  top=5pt,
  bottom=5pt,
  title=#1
}

\newtcolorbox{summarybox2}[1][]{
  colback=gray!10,
  colframe=gray!100,
  arc=1mm,
  boxrule=0.5pt,
  left=5pt,
  right=5pt,
  top=5pt,
  bottom=5pt,
  title=#1
}

\newmdenv[
  linewidth=1.5pt,
  linecolor=black,
  topline=false,
  bottomline=false,
  rightline=false,
  leftmargin=0pt,
  innertopmargin=0pt,
  innerbottommargin=0pt,
  innerleftmargin=5pt,
  innerrightmargin=0pt
]{myquote}

\newcommand{\quoteFont}[1]

\usepackage{tcolorbox}
\usepackage{mdframed}
 \usepackage{natbib}
 \usepackage{float}
 \usepackage{hyperref}

\newcommand{\imp}[1]{\textcolor{black}{#1}}
\newcommand{\minor}[1]{\textcolor{black}{#1}} 
\usepackage{longtable}
\usepackage{graphicx} 
\usepackage{tabularx}
\usepackage{booktabs}
\usepackage{url}
\usepackage{xspace}
\usepackage[table]{xcolor}
\usepackage{amsmath}
\usepackage{caption}
\usepackage{multirow}
\usepackage{makecell}
\usepackage{mdframed}
\usepackage{geometry}  
\usepackage{ragged2e}

\begin{document}

\title{Practitioner Insights on Fairness Requirements in the AI Development Life Cycle: An Interview Study}




\author{
Chaima Boufaied\\
University of Calgary\\
\texttt{chaima.boufaied@ucalgary.ca}
\and
Thanh Nguyen\\
University of Calgary\\
\texttt{thanh.nguyen2@ucalgary.ca}
\and
Ronnie De Souza Santos\\
University of Calgary\\
\texttt{ronnie.desouzasantos@ucalgary.ca}
}

\maketitle

\begin{abstract}
Nowadays, Artificial Intelligence (AI), particularly Machine Learning (ML) and Large Language Models (LLMs), is widely applied across various contexts. However, the corresponding models often operate as black boxes, leading them to unintentionally act unfairly towards different demographic groups. This has led to a growing focus on fairness in AI software recently, alongside the traditional focus on the effectiveness of AI models.
Through \total semi-structured interviews with practitioners from different application domains and with varied backgrounds across \countries countries, we conducted research on fairness requirements in AI from software engineering perspective. 
Our study assesses the participants’ awareness of fairness in AI / ML software and its application within the Software Development Life Cycle (SDLC), from translating fairness concerns into requirements to assessing their arising early in the SDLC. It also examines fairness through the key assessment dimensions of implementation, validation, evaluation, and how it is balanced with trade-offs involving other priorities, such as addressing all the software functionalities and meeting critical delivery deadlines.
Findings of our thematic qualitative analysis show that while our participants recognize the aforementioned AI fairness dimensions, practices are inconsistent, and fairness is often deprioritized with noticeable knowledge gaps. 
\imp{To further structure these insights, we modeled the key challenges in a framework that systematically organizes the findings and illustrates the main obstacles practitioners face when incorporating fairness in AI/ML projects.
These insights are reflected across participants, spanning different roles, seniority levels, and countries in which the AI projects were conducted.}
This highlights the need for agreement with relevant stakeholders on well-defined, contextually appropriate fairness definitions, the corresponding evaluation metrics, and formalized processes to better integrate fairness into AI/ML projects.
\end{abstract}
\noindent\hangindent=1.5em\hangafter=0\textbf{Keywords:} AI/ML fairness, fairness requirements, interview, qualitative analysis, \newline thematic analysis

\section{Introduction}\label{intro}

The widespread application of AI across various domains nowadays requires more than simply reporting a high effectiveness of the corresponding models, achieving, for instance, a high \emph{F1-score}. It also requires attention to the fairness of AI decisions, which are no longer an afterthought but a necessity in AI-based software~\cite{baresi2023understanding,chen2024fairness,singh2022fair, jakobssonrequirement}.

AI Fairness is typically defined as no discrimination against individuals or groups~\cite{chen2024fairness,ferrara2024fairness,lu2022towards}, due to their demographic attributes\footnote{We use attributes and features interchangeably throughout the paper.} such as race, religion, gender, sexual orientation, marital status, and social status (e.g., prioritization or favoritism of wealthy people over others in AI decision-making process)~\cite{cheng2021socially}.
While such attributes are commonly recognized as sensitive because they can unfairly influence AI decisions, leading to biased outcomes (i.e., misalignment between desired fairness conditions and actual outcomes, a.k.a., fairness bugs)~\cite{chen2024fairness}, their impact can vary depending on the context~\cite{demirchyan2025algorithmic,ferrara2024fairness,pant2025navigating,pham2025fairness,ramadan2025towards,ryanfairness,voria2024catalog}.
Non-demographic attributes, however, are generally not considered sensitive, as the corresponding AI decisions do not inherently result in biased outcomes.
For instance, in AI-based loan approval systems, race and gender attributes are sensitive as they relate to personal, protected characteristics that could lead the model to favor or disfavor certain groups (e.g., approving loans for white male applicants at higher rates than for others). In contrast, attributes such as credit score and annual income are viewed as legitimate, non-protected attributes~\cite{baresi2023understanding,pham2025fairness} because they reflect objective and task-relevant information.
Therefore, the desired fairness conditions in AI software represent \emph{fairness requirements}, whose satisfaction leads to fair, non–sensitive-attribute-based AI decisions, ensuring no discrimination towards individuals and groups~\cite{saxena2020fairness,ramadan2025towards,baresi2023understanding}.
Conversely, unmet fairness requirements cause harmful cycles of bias and social inequality, affecting various contexts, from hiring, loan approvals, and finance, to more critical application domains such as criminal justice and healthcare~\cite{Makhlouf2024,gunasekara2025systematic,yang2024survey,baresi2023understanding,holstein2019improving,ferrara2024fairness,deng2022exploring,van2020ethical,voria2024catalog,ramadan2025towards,ryanfairness,cheng2021socially,bahangulu2025algorithmic}.
Several definitions of fairness exist in AI/ML systems~\cite{mehrabi2021survey,baresi2023understanding,cheng2021socially,demirchyan2025algorithmic,deng2022exploring,ferrara2024fairness,ferrara2024refair,pant2025navigating,stoyanovich2023fairness,tang2025towards,pham2025fairness,ryan2023integrating,ramadan2025towards}, including statistical/demographic parity (similar outcomes across groups), equalized odds (i.e., achieving similar true positive and false positive rates, TPR and FPR, respectively), fairness through unawareness (avoiding bias by not feeding sensitive attributes to AI models), and fairness through awareness (treating similar individuals equitably in the presence of sensitive attributes). Fairness requirements are derived from these definitions and address sources of unfairness such as i) data bias (label, feature, or distribution)~\cite{mehrabi2021survey,ferrara2024fairness,pham2025fairness,ryan2023integrating,ramadan2025towards,lu2022towards,cheng2021socially}, ii) algorithmic bias (post-data processing)~\cite{ferrara2024fairness,holstein2019improving,pham2025fairness}, iii) user interaction bias, and human-driven discrimination~\cite{mehrabi2021survey,pham2025fairness,ferrara2024fairness,holstein2019improving,ryan2023integrating,ryanfairness}.

The diversity and conflict of fairness definitions also impact the choice of fairness metrics~\cite{van2020ethical,pham2025fairness,ryan2023integrating,voria2024catalog}. This requires an investigation of the operationalization of fairness in AI projects, including i) how fairness is defined and translated into actionable requirements, ii) how fairness concerns arise early in the SDLC process, and iii) how these requirements are subsequently implemented, validated, and evaluated throughout the AI lifecycle.
This paper addresses this gap by investigating practitioners' perspectives on fairness in AI through \total semi-structured interviews with software practitioners across different organizations and countries, with a focus on the following dimensions:
\begin{itemize}
    \item how AI fairness is perceived and defined in practice;
    \item how fairness requirements are translated from fairness concerns and managed (mainly documented and refined in light of discussions between involved stakeholders) in AI software;
    \item how fairness concerns arise early in the SDLC process, considering the corresponding challenges encountered;
    \item how fairness requirements are implemented, validated, evaluated, and traded off w.r.t., different technical, operational, and organizational constraints.
\end{itemize}

\imp{Through a thematic analysis of interview data collected from our \total participants, the four aforementioned dimensions provide a lens for understanding how fairness is handled across the AI/ML software lifecycle by software practitioners across various roles, expertise levels, and countries. 
These dimensions are not isolated but reflect a coherent progression across the AI/ML software lifecycle, from an abstract understanding of fairness to  its concrete application in AI software.
By looking at practitioners' understanding of AI fairness, how fairness concerns are translated to fairness requirements, the corresponding challenges that appear early in the SDLC, and how fairness is implemented and balanced with other priorities, we can see patterns, obstacles, and practical approaches that happen in real projects.}

\paragraph{\imp{Main Findings Summary.}}
\imp{Our interviews revealed several key insights about fairness in AI/ML systems:}
\imp{\begin{itemize}
    \item Our practitioners from different organizations across various countries with different roles and expertise levels showed nuanced understanding of fairness, moving from initial confusion with AI/ML accuracy, along with ambiguity regarding fairness concept and the corresponding metrics to recognizing multiple fairness dimensions such as data quality, model behavior, sensitive attributes, and ethical accountability.
    \item Fairness requirements were defined and addressed through a combination of technical, managerial, and operational practices, including data visualization, balancing, cleaning, and augmentation. However, these practices were applied inconsistently, with observed gaps at the level of the documentation and handling of fairness requirements in the software development process, and stakeholder-related interpretation of such requirements.
    \item Fairness concerns arise early in the development lifecycle, often due to data representation bias, imbalance and labeling issues, and data coverage gaps. Additionally, various challenges in identifying and applying fairness in AI software were reported, such as data and model challenges, unclear fairness definitions, limited metric support, and tight time constraints.
    \item Fairness promotion techniques varied widely across projects, where fairness validation was often challenging and inconsistent. Plus, fairness was frequently deprioritized relative to model performance, deadlines, and missing functional features, highlighting the need for clearer policies and structured evaluation.
\end{itemize}
}

\emph{Paper Structure.} The remainder of this paper is structured as follows: Section~\ref{backg} introduces relevant background concepts. Section~\ref{related} reviews related work. Section~\ref{method} outlines the
research methodology used for our qualitative analysis. Section~\ref{eval} presents the findings of our research questions, based on the responses of interview participants. 
Section~\ref{disc} \imp{presents insights from our findings, with a mapping to prior research, highlighting areas of alignment where applicable, followed by the corresponding practical implications and recommendations. It also proposes a key challenges framework in which we systematically categorize the main fairness issues shared by our participants, and addresses threats to validity and limitations of our study}.
Finally, Section~\ref{conc} concludes the paper, providing directions for future work.
\section{Background}\label{backg}
The increasing use of AI nowadays across various domains, notably criminal justice, finance, and healthcare~\cite{voria2024catalog, yang2024survey, gunasekara2025systematic,ryanfairness,cheng2021socially,ramadan2025towards,bahangulu2025algorithmic,Makhlouf2024,baresi2023understanding,holstein2019improving,ferrara2024fairness,deng2022exploring,van2020ethical} has raised the need for assessing the fairness of AI decisions towards different demographic groups, w.r.t., specific characteristics such as race, gender, religion, and age. 
These characteristics, known as sensitive attributes (also called protected attributes~\cite{chen2024fairness,baresi2023understanding,cheng2021socially,yang2024survey,mujtaba2025behind}), may 
introduce bias in AI decisions (\imp{i.e., biased AI model behavior}), causing the AI system to treat individuals or groups unequally~\cite{chen2024fairness}.

\imp{Discrimination based on gender and race is among the most commonly reported forms of unfair treatment in AI software in various contexts ~\cite{bahangulu2025algorithmic,mujtaba2025behind,fabris2025fairness,pant2025navigating,pham2025fairness,ryan2023integrating,ryanfairness,sorathiya2024ethical,ramadan2025towards,biswas2024hi}. 
For example, an AI-powered recruitment tool for Amazon was found to systematically discriminate against female job applicants~\cite{bahangulu2025algorithmic,mujtaba2025behind}.
Additionally, a machine bias study in ProPublica\footnote{https://www.propublica.org/article/machine-bias-risk-assessments-in-criminal-sentencing}  demonstrates that risk assessment algorithms, such as COMPAS recidivism algorithm~\cite{cheng2021socially}, systematically overestimate recidivism risk for Black defendants compared to the white ones.
More critically, from a health-impact perspective, higher accuracy is reported for light-skinned than dark-skinned patients in diabetic retinopathy~\cite{yang2024survey}, leading to incorrect treatment recommendations.}

AI fairness definitions differ from a context to another~\cite{demirchyan2025algorithmic,ferrara2024fairness,pant2025navigating,pham2025fairness,ramadan2025towards,ryanfairness,voria2024catalog} but mainly framed around i) ensuring equal positive outcomes across groups, ii) treating similar
inputs consistently, and/or iii) preventing unintended causal relationships between features and predictions~\cite{voria2024catalog}.
Further, attempting to adopt multiple fairness definitions simultaneously can result in contradictions with the overall project objectives, making it difficult to satisfy different fairness metrics at the same time (a.k.a., the impossibility theorem~\cite{baresi2023understanding}). For instance, enforcing positive predictive value parity between Black and White defendants in COMPAS~\cite{cheng2021socially} prevents the equalization of error rates. Similarly, achieving both demographic parity and statistical parity is highly unlikely, as while the former assesses the likelihood of a positive outcome, the latter evaluates the AI decision accuracy~\cite{demirchyan2025algorithmic,ramadan2025towards,ferrara2024fairness,ryan2023integrating,ryanfairness,chen2024fairness,deng2023investigating}. Thus, achieving agreement on which fairness definitions to adopt is essential before going further with the AI-based software development~\cite{ryan2023integrating,ferrara2024fairness}.
More specifically, it is recommended to adopt context-aware fairness definitions and strategies~\cite{mccormack2024ethical,voria2024catalog,ferrara2024refair,ferrara2024fairness} in software projects, while reaching consensus among all relevant stakeholders (e.g, developers,  policymakers, AI / ML experts, and lawyers)~\cite{singh2022fair,lu2022towards,ferrara2024fairness,mccormack2024ethical,demirchyan2025algorithmic,pant2025navigating,pham2025fairness, ramadan2025towards,ryan2023integrating,voria2024catalog} early in the SDLC, and providing a foundation for selecting and applying appropriate fairness metrics~\cite{ramadan2025towards,chen2024fairness,ryan2023integrating}.
Although AI explainability is widely seen as essential for transparency, accountability, and building trust~\cite{ferrara2024fairness,cheng2021socially,bahangulu2025algorithmic,gunasekara2025systematic,mccormack2024ethical,lu2022towards,xu2025shaping}, 
\imp{it is also closely connected to fairness, as the corresponding definitions are often linked to model interpretability and explainable AI principles~\cite{pawar2020explainable}.}

In AI systems, fairness requirements are conditions that are meant to ensure that AI decisions are not biased towards any individuals or groups, treating all equitably~\cite{saxena2020fairness,ramadan2025towards,baresi2023understanding}. However, these requirements are often vague in practice. For instance, the fairness requirement \emph{FR1:} ``The loan approval system must ensure that applicants are treated fairly regardless of their financial background.''~\cite{ramadan2025towards} is imprecise, leaving developers unsure how to measure or enforce fairness in AI software. The challenge is compounded by the multitude of fairness definitions and evaluation metrics from both individual and group perspectives~\cite{chen2024fairness,baresi2023understanding,holstein2019improving,pham2025fairness,ramadan2025towards,cheng2021socially,ferrara2024fairness,mujtaba2025behind,smith2025pragmatic}, with little guidance on which to adopt. More in detail, individual fairness (e.g., fairness through awareness/unawareness, counterfactual fairness, causal fairness)
~\cite{chen2024fairness,mujtaba2025behind} requires that similar individuals are treated similarly, such as two equally qualified and experienced employees with different genders receive similar chances from an AI system to get promoted. In contrast, group fairness (e.g., statistical parity, equalized odds, equal opportunity) emphasizes achieving equitable outcomes across demographic groups~\cite{chen2024fairness,baresi2023understanding,ryan2023integrating,mujtaba2025behind}, such as ensuring equal loan approval and denial rates across women and men with same financial situations. 
As a result, translating abstract fairness concerns into actionable software requirements is challenging~\cite{deng2023investigating,ferrara2024fairnessinterview,holstein2019improving,de2025software,ryan2023integrating,ryanfairness,lu2022towards,pant2025navigating}, making it crucial to \imp{understand how fairness considerations can be interpreted and operationalized within AI systems.}

Understanding the above AI fairness dimensions provides the foundation for addressing key challenges in defining, managing, and integrating AI fairness in the SDLC, while also assessing its importance relative to other priorities such as meeting strict software delivery deadlines, ensuring all promised functionalities, and maintaining high effectiveness.

\section{Related Work}\label{related}

\minor{In this section, we review the landscape of AI fairness research, covering i) key challenges and trade-offs, ii) fairness-aware development practices across the SDLC, and iii) limitations of existing empirical studies that further motivate the need for our work.}
\subsection{\minor{Fairness Challenges in AI Development}}
\imp{Prior research has documented numerous instances of unfair treatment in AI-supported systems across different application domains (see Section~\ref{backg} for different biased outcome observed in AI software).}
\imp{Biased AI treatment often stems from the violation of core fairness requirements embedded in the AI system design.
Among factors that lead to compromising fairness requirements, we mention biased or unrepresentative datasets~\cite{nguyen2025gray,van2020ethical,cheng2021socially,baresi2023understanding,joshi2021ai,yang2024survey,sorathiya2024ethical,jakobssonrequirement,holstein2019improving,de2025software,fabris2025fairness}, algorithmic design~\cite{de2025software,fabris2025fairness}, and limited transparency in data provenance~\cite{pushkarna2022data}. Human and organizational issues such as limited diversity, insufficient resources, and lack of accountability also contribute to inconsistent or ineffective fairness practices~\cite{ferrara2024fairness,deng2023investigating,rakova2021responsible,holstein2019improving,de2025software,banks2025multiple}. 
These \imp{unmet} fairness requirements can undermine trust in AI systems and lead to serious ethical and social consequences (e.g., stereotype reinforcement, reputational damage)~\cite{nguyen2025gray,baresi2023understanding, gunasekara2025systematic, sorathiya2024ethical, jakobssonrequirement, holstein2019improving, deng2023investigating, singh2022fair, ferrara2024fairness, ramadan2025towards, rakova2021responsible, ryan2023integrating, ryanfairness, tang2025towards, pant2025navigating}, making it crucial to examine how fairness concerns are identified, addressed, and managed throughout the SDLC.
These fairness requirement violation causes and the corresponding consequences are further corroborated by a recent gray literature investigation~\cite{nguyen2025gray}.
}

Some studies report a misalignment between fair ML research and industry practice from real-world perspectives~\cite{rakova2021responsible,holstein2019improving,mccormack2024ethical,deng2023investigating,lu2022towards,shin2019role}. 
For instance, \citet{rakova2021responsible} highlighted a tension between academic research and industry practice in communities like FAccT (Fairness, Accountability, and Transparency) where practical methods may meet disapproval from researchers, and industry successes may not align with research norms, emphasizing the need to understand these dynamics in ML-driven systems. Similarly, ethical-aware algorithms are sometimes too complicated to explain to stakeholders with no strong mathematical background and/or little connection to the AI development process~\cite{lu2022towards}.
Further, using semi-structured interviews and an anonymous survey of 267 ML practitioners, \citet{holstein2019improving} found that while research emphasizes algorithmic de-biasing, quantitative metrics, high-stakes domains, and often overlooks human or organizational biases, practitioners prioritize data collection and auditing with limited demographics, making it challenging to apply research metrics in practice.
Additionally, a recent SLR~\cite{sorathiya2024ethical} identifies two main gaps in 47 studies on extracting ethical concerns from user reviews: 1) unreliable methods for detecting fairness and transparency issues\footnote{Although fairness and transparency may sound similar, the former focuses on preventing discrimination and ensuring equitable treatment, while the latter focuses on openness and disclosure needed to build public trust.}, and 2) a lack of approaches to translate ethical concerns into concrete requirements in the requirement elicitation phase of the SDLC.

\minor{Further,} the focus on AI fairness may come with various tradeoffs. This explains why teams are often recommended to carefully balance their different objectives, as later interventions may be costly or resource-intensive~\cite{holstein2019improving,rakova2021responsible,pushkarna2022data,baresi2023understanding,demirchyan2025algorithmic,ferrara2024fairness}. For example, efforts to improve AI fairness through modifications to models and datasets enhanced fairness according to quantitative metrics but unintentionally harmed user experience and software functionality, affecting functional requirements~\cite{holstein2019improving,lu2022towards,pant2025navigating,cheng2021socially,xu2025shaping,smith2025pragmatic}. 
Additionally, organizational goals such as meeting strict deadlines often conflict with fairness, where business objectives always receive maximum attention, while considerations of fairness are either treated as optional or addressed only when time allows~\cite{rakova2021responsible, pushkarna2022data,pant2025navigating,ryan2023integrating,holstein2019improving,deng2023investigating,ferrara2024fairnessinterview}.
Additional fairness trade-offs include a balance between: i) providing detailed fairness information while keeping AI documentation understandable~\cite{pushkarna2022data}, ii) ensuring security such as protecting data privacy so that sensitive information is not shared~\cite{pushkarna2022data,lu2022towards,tang2025towards,mccormack2024ethical}, and iii) guaranteeing a high effectiveness (referred to inconsistently and interchangeably in the literature as accuracy, performance, and predictive accuracy~\cite{demirchyan2025algorithmic,ferrara2024fairness,ferrara2024refair,holstein2019improving, pham2025fairness, ryan2023integrating,ryanfairness,chen2024fairness,voria2024catalog,pant2025navigating,de2025software,cheng2021socially,tang2025towards,mccormack2024ethical,smith2025pragmatic,ferrara2024fairnessinterview,rakova2021responsible,deng2023investigating}).

\subsection{\minor{Fairness Practices Across the SDLC}}

Given the lack of awareness and understanding of fairness as well as the absence of clear guidelines and tools to apply it throughout the AI development lifecycle~\cite{voria2024catalog,ferrara2024fairnessinterview,de2025software,pant2025navigating,shin2019role}, many studies recommend the integration of fairness considerations throughout the SDLC~\cite{singh2022fair,voria2024catalog,ferrara2024refair,lu2022towards,ferrara2024fairness,ramadan2025towards,ryan2023integrating,chen2024fairness,holstein2019improving,pham2025fairness,deng2023investigating,ferrara2024fairnessinterview}.
Therefore, various fairness-aware practices have been mapped into different stages through strategies that identify and mitigate biases across the AI pipeline~\cite{ferrara2024fairness}, with fairness treated as a first-class entity from early development stages~\cite{ramadan2025towards,chen2024fairness}.
These fairness practices are categorized according to the ML pipeline in a recent study~\cite{ferrara2024fairnessinterview} as follows:
i) pre-processing through fair data selection, collection, and preparation (e.g., data representation, balancing, rule-based scraping to remove sensitive attribute information from features, rule-based substitution by neutralizing sensitive keywords)~\cite{ferrara2024fairnessinterview,fabris2025fairness,lu2022towards,de2025software}, 
ii) in-processing via fairness-aware model design~\cite{fabris2025fairness,ferrara2024fairnessinterview,lu2022towards}, and iii) post-processing through equitable evaluation and testing~\cite{de2025software,fabris2025fairness,ferrara2024fairnessinterview,lu2022towards}. These align with responsible AI-by-design practices~\cite{lu2022towards,ferrara2024fairnessinterview,de2025software,fabris2025fairness}, spanning from ethical requirement specification and value-sensitive design to ethical coding, testing, and model deployment \cite{ferrara2024fairnessinterview,lu2022towards,ryan2023integrating,ryanfairness}.
Building on these approaches, \citet{singh2022fair} proposed the Fair CRISP-DM model, recommending formal plans to address fairness such as screening variables, planning audits, considering human supervision, and establishing an appeals process while actively mitigating data biases (e.g., historical, representation, sampling) during data understanding and preparation. The model also incorporates fairness directly during the model training phase using techniques like constrained optimization and fairness penalties in the loss function, rather than post-training fixes. Overall, delivering fair and ethical AI software is a nuanced goal that needs to be addressed throughout the entire SDLC process~\cite{ryan2023integrating}.

\subsection{\imp{Limitations of Existing Empirical Studies}}
\imp{Several empirical studies on AI fairness have been proposed in the literature, employing either} i) quantitative analyses by means of large-scale surveys~\cite{ferrara2024fairnessinterview}, ii) qualitative analyses, mostly by means of semi-structured interviews~\cite{de2025software,pant2025navigating,rakova2021responsible,ryan2023integrating,smith2025pragmatic,ryanfairness}, or iii) mixed analyses, where semi-structured or in-depth interviews are combined with other techniques like workshops or quantitative  surveys~\cite{deng2023investigating,holstein2019improving,shin2019role} \imp{(see Table~\ref{empirical-limitations})}.
\imp{The qualitative data analysis techniques rely on different thematic analysis variants, such as reflexive thematic analysis~\cite{ryan2023integrating,smith2025pragmatic}, inductive or phase-based thematic analysis with open, axial, and selective coding~\cite{de2025software,deng2023investigating}, template analysis~\cite{ryanfairness}, affinity diagramming for bottom-up theme synthesis~\cite{holstein2019improving,rakova2021responsible}, and socio-technical grounded theory~\cite{pant2025navigating}.
However, methodological reporting varies: while most semi-structured interviews describe collaborative coding and theme development~\cite{de2025software, pant2025navigating, rakova2021responsible, ryan2023integrating, ryanfairness, smith2025pragmatic}, only one study~\cite{de2025software} explicitly reports the data saturation strategy. 
Further, we noticed a common incomplete reporting of interview protocols such as the recruitment process, interview duration, and/or the used platforms~\cite{de2025software, deng2023investigating, holstein2019improving, rakova2021responsible, ryanfairness, shin2019role, smith2025pragmatic}.
These variations highlight differences in methodological transparency and depth across prior qualitative studies.
}

\imp{In the following, we further critically review these prior empirical studies and highlight their limitations w.r.t., the following additional aspects:
\begin{itemize}
    \item \textbf{Scope Coverage:} The SDLC phases covered by the study (i.e., requirements, design, implementation, testing, deployment, maintenance), as inferred from the interview guide questions.
    \item \textbf{Sample Coverage:} Participants diversity in terms of countries in which the AI/ ML projects were conducted, their organizational roles, their experience level (seniority) coverage, and the number of participants of the study.
    \item \textbf{Methodological Coverage:} We report the empirical study type (qualitative, quantitative, or mixed), along with the exact study type (e.g., semi-structured interview, structured interview, survey, or workshop).
\end{itemize}
}
\imp{\begin{longtable}{p{.5cm} p{2cm} p{1cm} p{2.7cm} p{1.5cm} p{.7cm} p{2.85cm}}
\caption{\imp{Summary of prior empirical studies and their limitations.}}
\label{empirical-limitations} \\
\toprule
\textbf{Study} & \centering \textbf{Scope \newline Coverage} & \multicolumn{4}{c}{\textbf{Sample Coverage}} & \textbf{Method. Coverage} \\
\cmidrule(lr){3-6}
 &  & \textbf{Country} & \centering \textbf{Role} & \textbf{Seniority} & \textbf{Size} &  \\
\midrule
\endfirsthead

\endhead
\endfoot
\bottomrule
\endlastfoot \rowcolor{gray!10}
\cite{de2025software} & Development, Testing & \centering 1 & AI/ML, Software Engineering, \newline Research, \newline Managerial & Junior, \newline Mid-level, \newline Senior, \newline Highly \newline senior & 22 & Qualitative \newline (semi-structured \newline interview) \\ 
\cite{deng2023investigating} & Requirement, \newline Design, \newline Development, \newline Testing & \centering 7 & AI/ML, Software Engineering, \text{UX, Managerial,} Research & N/A & 17 & Qualitative \newline (semi-structured \newline interview) \\ 
& & & & & 12 & Qualitative \newline \text{(design workshop)} \\
\rowcolor{gray!10}
\cite{ferrara2024fairnessinterview} & Requirement, \newline Design, \newline Development & \centering N/A & Software \text{Engineering,} \text{AI/ML, Managerial} & Junior, \newline Senior & 117 & Quantitative \newline (survey) \\
\cite{holstein2019improving} & Design, \newline Development, \newline Design, \newline Testing & \centering N/A & AI/ML, Managerial, UX, Software Engineering, Research & N/A & 35 & quantitative \newline (semi-structured \newline interview) \\ 
& & & & & 267 & Qualitative \newline (survey) \\ \rowcolor{gray!10}
\cite{pant2025navigating} & Requirement, \newline Testing, \newline Deployment & \centering 8 & \text{AI/ML, Software} Engineering & Junior, \newline Senior $\circ$ & 22 & Qualitative \newline (semi-structured \newline interview)\\
\cite{rakova2021responsible} & Design
 & \centering 5 & Software \newline Engineering, \text{Research,} \newline Organizational, Managerial & Junior, \newline Senior $\circ$ & 26 & Qualitative \newline (semi-structured \newline interview) \\ \rowcolor{gray!10}
\cite{ryan2023integrating} & Design, \newline Development, \newline Testing  & \centering N/A & \text{AI/ML, Software} Engineering, \newline Research, UX & N/A $\star$ & 18 & Qualitative \newline (semi-structured \newline interview) \\  
\cite{ryanfairness} & Design, \newline Development, \newline Testing & \centering 9 & Software \newline Engineering, \text{Managerial, UX} & N/A $\star$ & 16 & Qualitative \newline (semi-structured \newline interview) \\ \rowcolor{gray!10}
\cite{shin2019role} & Design, \newline Development & \centering N/A & \centering N/A & N/A $\star$ & \text{$\geq 230$} & Quantitative \newline (survey) \\ 
\rowcolor{gray!10}
&&&&& N/A & \text{Qualitative (in-depth} \newline interview) \\
\cite{smith2025pragmatic} & Testing & \centering N/A$\bullet$ & \text{AI/ML, Software} Engineering, UX & Senior & 21 & Qualitative \newline (semi-structured \newline interview) \\ \rowcolor{gray!10}
\textbf{Us} & \textbf{Requirement, \newline 
Design, \newline Development, Testing} & \centering \textbf{23} & \textbf{AI/ML}, \newline \textbf{Research}, \newline \textbf{Software \newline Engineering}, \newline \textbf{Managerial} & \textbf{Junior}, \newline \textbf{\text{Mid-level}}, \newline \textbf{Senior}, \newline \textbf{Highly senior} & \textbf{26} & \textbf{Qualitative \newline (semi-structured \newline interview)} \\
\multicolumn{7}{l}{}\\
\multicolumn{7}{l}{\footnotesize{\shortstack[l]{$\circ$ No explicit seniority levels are defined, though junior ($\le$2 years) and senior (
$\ge$5 years) can be inferred.}}} \\
\multicolumn{7}{l}{\footnotesize{\shortstack[l]{$\star$ Either only age ranges of participants were shared, which does not necessarily reflect the overall \\ seniority level, or senior titles were reported for some participants only.}}} \\
\multicolumn{7}{l}{\footnotesize{\shortstack[l]{$\bullet$ Two countries are explicitly mentioned (Philippines and Austria);
the rest refer to regions or ethnicities.}}} \\
\multicolumn{7}{l}{\textbf{Role Reference:}}\\
\multicolumn{7}{l}{\footnotesize{\textbf{AI/ML}: e.g., ML engineer, AI engineer, computer vision engineer}}\\
\multicolumn{7}{l}{\footnotesize{\shortstack[l]{\textbf{Software Engineering}: e.g., software engineer / architect / analyst, programmer, tester, product designer}}}\\
\multicolumn{7}{l}{\footnotesize{\textbf{Managerial}: e.g., project / product / data science manager, CEO}}\\
\multicolumn{7}{l}{\footnotesize{\shortstack[l]{\textbf{Research}: e.g., research scientist / software engineer, research fellow,
R\&D specialist}}}\\
\multicolumn{7}{l}{\footnotesize{\textbf{UX}: e.g., user experience practitioner, Human-Computer Interaction (HCI)}} \\
\multicolumn{7}{l}{\footnotesize{\textbf{Organizational}: Non software-related roles, such as HR specialist, legal counsel, and marketing / sales}} \\
\end{longtable}
}

\imp{As shown in Table~\ref{empirical-limitations}, most of the prior empirical studies, notably those relying solely on semi-structured interviews~\cite{de2025software,pant2025navigating,rakova2021responsible,ryan2023integrating,ryanfairness,smith2025pragmatic}, do not comprehensively cover all phases of the AI/ML software development lifecycle, with a primary focus on  design and/or development phases~\cite{deng2023investigating,holstein2019improving,ryan2023integrating,ryanfairness,shin2019role,ferrara2024fairnessinterview,rakova2021responsible}, and testing~\cite{deng2023investigating,de2025software,holstein2019improving,pant2025navigating,ryan2023integrating,ryanfairness,smith2025pragmatic}, with little attention given to the requirement elicitation phase~\cite{deng2022exploring,ferrara2024fairnessinterview,pant2025navigating}.
}
\imp{In terms of sample coverage, several studies either do not report country information (e.g., \cite{ferrara2024fairnessinterview, holstein2019improving, ryan2023integrating, shin2019role, smith2025pragmatic}) or focus on a limited set of countries (from one country~\cite{de2025software} to nine countries~\cite{ryanfairness}). 
Role coverage often includes User Experience (UX) or organizational roles~\cite{deng2022exploring,holstein2019improving,ryan2023integrating,ryanfairness,smith2025pragmatic}, limiting focus on direct AI/ML and software engineering practitioners. 
Seniority level of participants is inconsistently reported: some studies do not define it~\cite{deng2023investigating, ryan2023integrating, ryanfairness, shin2019role,holstein2019improving,pant2025navigating,rakova2021responsible}), while others either only share participants age~\cite{ryan2023integrating,ryanfairness,shin2019role}, which does not necessarily reflect their years of experience, or are restricted to senior level only~\cite{smith2025pragmatic}.
In light of these scope and sampling coverage limitations of prior empirical studies, our work is the first to systematically cover the full AI/ML lifecycle while including diverse roles, well-defined seniority levels, and broad geographic representation spanning across 23 countries, providing a comprehensive view of industry practices. 
}


\imp{Beyond scope and sample characteristics,} while these prior empirical studies have:
i) advanced our understanding of how fairness is interpreted~\cite{ferrara2024fairnessinterview, pant2025navigating, ryan2023integrating} (especially in terms of clarity, usefulness, and feasibility~\cite{ferrara2024fairnessinterview}) and evaluated~\cite{ferrara2024fairnessinterview, ryan2023integrating, ryanfairness, de2025software, pant2025navigating, smith2025pragmatic};
ii) examined how fairness is perceived and traded off against other aspects such as AI model effectiveness, security, meeting strict deadlines, supporting missing functional or non-functional requirements, user satisfaction, and stakeholder involvement~\cite{de2025software, ferrara2024fairnessinterview, pant2025navigating, smith2025pragmatic, deng2023investigating, holstein2019improving, rakova2021responsible, ryan2023integrating, ryanfairness};
iii) identified where in the ML pipeline fairness concerns typically emerge, along with the corresponding challenges~\cite{ferrara2024fairnessinterview, holstein2019improving, ryan2023integrating}; and
iv) focused on testing fairness by means of the proper strategies and tools~\cite{ferrara2024fairnessinterview, de2025software, holstein2019improving, pant2025navigating},
none provided a comprehensive, structured analysis of how fairness concerns are formally translated into fairness requirements, officially documented, and validated throughout the entire SDLC process from a software engineering perspective.
\imp{This paves the way for our semi-structured interview study to assess how practitioners 1) perceive and understand key fairness concepts and how fairness has been reflected in their work, 2) identify, interpret, document, and refine fairness requirements in practice, with a focus on how stakeholders are involved in their interpretation, 3) investigate how fairness concerns arise in the early stages of the SDLC and the corresponding challenges, and 4) implement, validate, and evaluate AI fairness, as well as manage fairness trade-offs with other project goals such as model performance, usability, deadlines, and functionality prioritization.}

\section{Methodology}\label{method}
We conducted a qualitative study to investigate how AI/ML practitioners with different roles, across \countries countries, handle fairness requirements during the SDLC. This section details our research methodology, which is mainly based on three phases:
\begin{itemize}
    \item \emph{Pre-interview Phase}: Design the interview guide, recruit participants, prepare logistics, and obtain a formal consent, carefully read, signed, and sent back by all the participants prior to interviewing them.
    \item \emph{Interview Phase}: Conduct pilot and individual interviews.
    \item \emph{Post-interview Phase}: Collect responses, perform thematic analysis, and report actionable findings.
\end{itemize}

\subsection{Research Questions}\label{rqs}
With the exponentially use of AI/ML systems nowadays, ensuring fairness of such systems is increasingly critical, particularly when targeting sensitive domains such as hiring, and crime prediction. Beyond understanding AI/ML model fairness, it is essential to explore how fairness requirements are managed (i.e., identified, interpreted, refined, and documented) throughout the SDLC. 
Therefore, our study aims to: i) examine practitioners’ understanding of fairness in AI/ML; ii) investigate how fairness-related requirements are handled during the software development process, iii) explore how fairness concerns arise  early in the SDLC, with a focus on the corresponding challenges, and iv) understand how fairness requirements are implemented, validated, and evaluated in practice, and how practitioners handle fairness trade-offs with other project goals. 
To gain these insights, we address the following research questions:

\begin{itemize}
    \item \textbf{RQ1:} How aware are AI/ML professionals of key concepts of fairness in AI/ML software development, and how fairness has been reflected in their work?
    \item \textbf{RQ2:} How are fairness concerns translated into fairness requirements? And how are the latter refined, documented, and interpreted by stakeholders in AI/ML software development process?
    \item \textbf{RQ3:} How do fairness concerns arise and manifest early in AI/ML projects within the SDLC? And what challenges do teams face in handling them?
    \item  \textbf{RQ4:} How do teams implement, validate, and evaluate fairness requirements during development?  And how do they manage trade-offs with other project goals?
\end{itemize}

\subsection{Interview Guide Design}\label{guide}
We designed an interview guide comprising 15 open-ended questions.
The guide begins with introductory questions covering gender, job title, education, application domain, main responsibilities, experience, and country to contextualize the participants’ roles and their work environments, \imp{followed by sub-questions organized by research question to maintain a consistent and logical structure
(the full list of questions is provided in the interview guide in appendix~\ref{appendix})}.
Formally, each of the aforementioned research questions $RQi$ (i = 1..4) is addressed through $x$ sub-questions ($Qj$, where $j = 1..x$), denoted as $RQi_{Qj}$. 
\imp{The order of our interview sub-questions was maintained for all participants, ensuring consistent sequencing. Each sub-question included predefined cues to clarify or expand responses as needed}. 

For instance, to answer RQ1, we asked the following two open-ended sub-questions:
\begin{itemize}
    \item \textbf{$RQ1_{Q1}$:} When you hear the term fairness in the context of software or AI development, what does it mean to you?
    \item \textbf{$RQ1_{Q2}$:} Can you share an example from your work where software fairness was a concern or topic of discussion during the software development process?
\end{itemize}
While $RQ1_{Q1}$ assesses conceptual awareness by asking participants how they define fairness, $RQ1_{Q2}$ probes practical awareness through real-world examples. Together, these two questions explore practitioners’ understanding of AI/ML fairness and their hands-on experience with such systems, directly supporting RQ1.

\subsection{Participant Recruitment \imp{Protocol}}\label{protocol}
\imp{This section details our participant recruitment protocol, in which we mainly clarify i) the target population, with screening criteria and sampling strategy, and ii) the recruitment and interview procedures, including passive and active invitations, messaging, and interview scheduling.}

\subsubsection{\imp{Target Population}}\label{targetpopul}
\imp{
The participants in this study were selected to represent practitioners actively involved in AI/ML system development, deployment, evaluation, or management within software projects (screening criterion). Throughout the recruitment process, each potential participant was asked to confirm their use of AI/ML in the software systems they were involved in. We only considered individuals who confirmed hands-on or decision-level involvement in AI/ML projects. This confirmation was based on information shared through their professional profiles and descriptions of projects they had worked on. Individuals who indicated that their knowledge of AI/ML was theoretical, limited, or not directly tied to practical project involvement were excluded based on these inclusion/exclusion criteria (screening criterion). 
Therefore, interviews were scheduled only with those who demonstrated concrete experience in AI/ML systems (sampling strategy).
}
\begin{longtable}{p{.6cm} | p{2.5cm} | p{1.5cm} | p{1.3cm} | p{3cm} | p{1cm} | p{1.7cm}}
\caption{Participants Demographics; ID=Participant ID; Country [CA=Canada, LU=Luxembourg, PK=Pakistan, EG=Egypt, UK=United Kingdom, US=United States of America, TU= Tunisia, IR=Iran, EC=Ecuador, LK=Sri Lanka, SG=Singapore, SK= South Korea, NO=Norway, VN=Vietnam, DZ= Algeria, IN=India, DE=Germany, BD=Bangladesh, CN=China, AE=United Arab Emirates, NE=Nepal, MA=Morocco, BR=Brazil]}     \label{demographics}
\\
\rowcolor{gray!30}
\textbf{ID} & \textbf{Role} & \textbf{Edu. Bg} & \textbf{Gender} & \textbf{App. Domain} & \textbf{Exp (years)} & \textbf{Country} \\
\hline
\endfirsthead

\rowcolor{gray!30}
\textbf{ID} & \textbf{Role} & \textbf{Edu. Bg} & \textbf{Gender} & \textbf{App. Domain} & \centering \textbf{Exp (years)} & \textbf{Country} \\
\hline
\endhead
\rowcolors{2}{gray!10}{white} 
         P1 & \text{Data science} manager & Masters & \centering M & E-commerce, healthcare, social services & \centering 9 & US, CA \\ \rowcolor{gray!10}
         P2 & Data scientist & Masters & \centering M & Finance & \centering 4 & BR \\   
        P3 & Software project manager & Bachelor & \centering M & Web and mobile applications & \centering 2 & PK \\ \rowcolor{gray!10}
        P4 & \text{ML software} \text{developer} & Masters & \centering M & Computer vision & \centering 4 & CA \\
        P5 & AI engineer & Bachelor & \centering F & Telecommunication & \centering  5 & EC \\ \rowcolor{gray!10}
        P6 & Data scientist & Bachelor & \centering M & Telecommunication & \centering  \centering  10 & EC \\
        P7 & Software \text{engineer} & Bachelor & \centering M & Finance \& Baking &  \centering 6 & CA \\ \rowcolor{gray!10}
        P8  & Research fellow & Bachelor & \centering M & Wireless communication  & \centering 3 & LK, SG \\
        P9  & Computer vision engineer & Bachelor & \centering M & Optic detection  &\centering  4-5 & SK, NO \\ \rowcolor{gray!10}
        P10  & AI researcher & Masters & \centering M & Health care in mobile applications  &  \centering 1-2 & VN \\
        P11  & \text{Professor in} \text{Computer Science} & PhD & \centering M & Health care  & \centering 18 & EG, US \\ \rowcolor{gray!10}
        P12 & Computer vision engineer & Masters & \centering F & Geo-spatial Vision & \centering  .5 & DZ \\
        P13 & AI researcher & PhD & \centering F & Image-based vision & \centering 10 & TU, CA \\ \rowcolor{gray!10}
        P14 & AI researcher & Bachelor & \centering M & Remote sensing prediction & \centering 3 & PK \\
        P15 & AI consultant & Masters & \centering F & Marketing \& Real estate & \centering 2-3 & US, CA, UK, IN \\ \rowcolor{gray!10}
        P16 & AI researcher & Masters & \centering M & Computer vision & \centering  4 & DE, IN, US \\
        P17 & ML researcher & Masters & \centering M & Multimodal Emotion Detection & \centering 4 & BD, CN \\ \rowcolor{gray!10}
        P18 & AI engineer & Bachelor & \centering M & Domain-specific NLP & \centering 2-3 & \text{EG, AE, USA} \\
        P19 & ML engineer & Bachelor & \centering M & E-commerce, health care & \centering 3-4 & \text{US, NP, SG} \\ \rowcolor{gray!10}
        P20 & R\&D specialist & Masters & \centering M & Domain-agnostic computer vision & \centering 3 & DE, MA \\
        P21 & R\&D specialist & Masters & \centering F & Domain-agnostic chatbot NLP & \centering 1-2 & IR \\ \rowcolor{gray!10}
        P22 & AI researcher & PhD & \centering F & Smart city applications & \centering 6 & IN \\
        P23 & AI researcher & Bachelor & \centering M & Smart city applications & \centering 2 & PK \\ \rowcolor{gray!10}
        P24 & AI engineer & Bachelor & \centering F & banking, telecommunication & \centering 3 & LK \\
        P25 & Data scientist & Bachelor & \centering F & Telecommunication, & \centering 4-5 & EC \\ \rowcolor{gray!10}
        P26 & R\&D specialist & Bachelor & \centering F & Agriculture, Health care, & \centering 1 & BD \\
        \bottomrule
\end{longtable}

Table~\ref{demographics} shows \total participants from various roles and domains of AI/ML.
Nine out of the~\total participants were female (representing $\approx$ 35\% of all participants). \imp{This reflects a lower representation consistent with the overall gender imbalance in computer science and software engineering departments~\cite{cheryan2017some}, where most practitioners are men, despite the gradual increase in female participation.}
The majority of participants hold either a Bachelor or Masters degree (13 and 10 participants, respectively), with Bachelor being slightly more common; PhD holders (only three participants) are notably fewer.
\imp{Practitioners are based in \countries countries (Column `Country' indicates the location(s) in which the AI/ML projects discussed during the interview were conducted, where some participants reported multiple countries due to cross-national project involvement).}
\imp{While we did not systematically collect data on the overall company size or the organizational maturity, most participants work in teams of 4 to 12 members, corresponding to small (4--5 members) and medium-sized (6--12 members) teams across organizations in multiple countries.}
Participants experience (Column `Exp (years)') ranges from six months to 18 years, with an average of $\approx$ 4.5 years. Their experience in AI/ML models varies across levels, comprising six junior ($\le 2$ years), 14 mid-level (2.5--5 years), three senior (6–9 years), and three highly senior (10+ years) professionals. 
\imp{This range ensures representation across varying levels of expertise relevant to AI/ML software development.}

Among our \total participants, several held core AI/ML roles such as AI/ ML engineer (P5, P18, P19, P24), ML software developer (P4), AI / ML researcher (P10, P13, P14, P16, P17, P22, P23), and computer vision engineer (P9, P12), with experience in industry, academia, or applied research, \imp{all of whom actively apply AI/ML in industry settings, providing insights grounded in real-world practice rather than purely academic perspectives.}
Others, though their titles do not explicitly state AI/ML specialization, such as R\&D specialist (P20, P21, P26), data scientist (P2, P6, P25), software engineer (P7), research fellow (P8), professor in computer science (P11), and managerial roles such as data science manager (P1) and software project manager (P3), have hands-on experience ranging from designing and fine-tuning models to managing teams and providing feedback on AI/ML projects.
Grouping participants based on their experience with AI/ML models, and based on their expertise in the domain enables the extraction of meaningful insights and the identification of relevant patterns.

\subsubsection{Interview Procedure}
\imp{Our participants were recruited through a combination of five personal contacts on professional platforms such as Teams (active recruitment), as well as an open call posted on LinkedIn (passive recruitment). 
More in detail, approximately 102 potential participants initially expressed interest via LinkedIn. Many of these individuals (around 40\%) later declined or did not respond, often citing lack of time, reconsideration of their participation, or misunderstanding the voluntary and non-committal nature of the study, leaving the remaining participants, of whom only those who met the inclusion/exclusion criteria (see Section~\ref{targetpopul}) moved forward in the study.
The observed diversity in roles, seniority levels, and countries reflects both the intentional effort to ensure coverage across these characteristics and naturally emergent outcomes from the recruitment process.} 

\imp{Our recruitment messages emphasized participants’ experience with AI/ML projects, without explicitly mentioning fairness, to avoid biasing the sample toward those who were already interested in fairness issues. Interviews were scheduled after participants confirmed their availability, coordinating logistics to accommodate different time zones and language preferences, and ensuring that participation and responses remained voluntary and anonymous.
}

We conducted all interviews remotely via Google Meet, and Microsoft Teams.
The interviews lasted between 26 and 60 minutes, with an average duration of ~45 minutes. According to steps in the \emph{interview phase}, we conducted pilot interviews, based on which we slightly improved some questions and clarified previously ambiguous phrasing. 
Although most of the interviews were conducted in English, some
interviews were conducted in French, according to the language preferences of the participants and their communication needs, while preserving the same semantics of the asked questions for the sake of consistency in data collection.

\imp{For participants who were unfamiliar with fairness, we provided a concise explanation: fairness was defined as no discrimination against individuals or groups~\cite{chen2024fairness}, based on sensitive attributes such as race, religion, gender, sexual orientation, marital status, and social status, providing concrete examples of biased AI decisions (e.g., prioritization or favoritism of wealthy people over others in AI decision-making)~\cite{cheng2021socially,nguyen2025gray}.
\imp{For instance, when we asked participants to define fairness (first question in RQ1), P23 responded: ``fairness is when data is accurate. If data is fair, the model is fair in the pipeline'', which is somehow similar to many prior participants who fully answered the interview without issues and demonstrated an understanding of fairness. However, when asked to give an example of a fairness concern (second question in RQ1), P23 replied as follows: ``...in covid-19 dataset... if accuracy is greater than 80\%, then the model is fair'', showing misunderstanding regarding the distinction between AI/ML fairness metrics and conventional model accuracy indicators. At this point, we provided the definition and the above example on favoritism. After then, P23 was able to give richer, more informed answers aligned with the interview questions.}
}

Although we did not conduct formal member checks due to the voluntary nature of the task and other priorities \imp{(none of the \total participants were willing to be contacted again after the interview)}, \imp{we confirmed our understanding of participants' answers in real time by asking questions such as} `Am I correct?' or `Is this what you meant?'.

\subsection{Ethics and Consent}
For each participant, we emailed a formal consent form stating that participation was voluntary, anonymous, and that no sensitive information would be asked or revealed. The participants were informed that they had the full right to withdraw at any time or to stop the interview altogether. We also gave them the choice to be recorded or not. While most agreed to recording, some declined. Interviews were conducted after participants carefully read the consent forms, signed, and sent them back to us. 

\subsection{Data Saturation}
\imp{This study follows a Thematic Analysis (TA) approach~\cite{clarke2017thematic}, in which segments of participants’ responses (a.k.a \emph{quotes}) were labeled to reveal patterns (called \emph{codes}) that are further grouped into broader concepts named \emph{themes}. Details on the used TA are shared in  Section~\ref{TA}.}
The responses of the participants varied significantly, and some individuals demonstrated a deep knowledge of fairness concepts in theory while also showing limited understanding or confusion between fairness and accuracy metrics. Due to the heterogeneity in expertise and perspectives, saturation emerged at different moments depending on the thematic category. This pattern is consistent with prior work showing that saturation often varies by topic and may be achieved earlier for clearly bounded themes and later for more complex or heterogeneous ones \cite{guest2006many}.

\imp{In our study, data saturation was assessed at the level of individual codes rather than at the theme level to capture the full range of distinct concepts in the data before we aggregate them into broader themes. More in detail, this thematic pattern reflects the characterization of saturation in the literature as the point at which additional interviews no longer produce new information that required the addition of new codes to the codebook or meaningful changes to category properties~\cite{fusch2015we}.} 
Therefore, subsequent mentions fitting existing codes were not considered new information. 
\imp{Table~\ref{tab:saturation} illustrates a sample of major dimensions we covered for each RQ, showing the participant for whom data saturation was reached (Column `Sat. P\#'), along with notes on subsequent contributions.
\begin{table}[h!]
\centering
\caption{\imp{Example of code-level saturation points for RQ1--RQ4.}}
\label{tab:saturation}
\begin{tabular}{m{.5cm}|m{4.5cm}|m{1.05cm}|m{6.3cm}}
\hline
\textbf{\imp{RQ}} & \textbf{\imp{Question Context}} & \textbf{\imp{Sat. P\#}} & \textbf{\imp{Notes on subsequent contributions}} \\
\hline
RQ1 & Fairness definition & P19 & Quotes from P20--P26 fit under existing codes \\
RQ2 & Stakeholder involvement & P24 & No new information added by P25 and P26 \\
RQ3 & Challenges in applying fairness & P26 & Last participant's quote led to a new  code\\ 
RQ4 & Fairness promotion techniques & P12 & Later contributions after P12 aligned with existing codes \\
\hline
\end{tabular}
\end{table}}

Throughout this study, we observed that new insights ceased to appear at different stages, depends on each sub-question under each RQ. 
\imp{For instance, in fairness definition (RQ1), data saturation was reached at participant P19, where the latter's quote states that ``there should not be any bias from feature perspective, while including all features... also when removing one or many of sensitive features, still the model shall remain fair.'', emphasizing fairness regardless of altering or removing sensitive features, where the quotes shared by participants P20--P26 did not lead to new codes. For example, 
P21's quote aligns with the code generated for P10's quote, both fitting under code \emph{Model robustness to sensitive attribute influence in decisions (IC9)} (see Table~\ref{fairness_def}, along with the supplementary material for access to the different codes generated~\cite{methodology_Excelsheet}), as both emphasize fairness across all data categories and sensitivity to protected attributes.
Further, for stakeholder involvement (RQ2), data saturation was reached at P24, where none of P25 and P26 shared relevant information that could potentially fit under a new code.
}
\imp{In terms of reported fairness challenges (RQ3), data saturation was reached at the level of P26, who finally shared a new fairness challenge that comes from AI-generated data, quoting the following: ``AI-generated images were a big concern; we didn't know that at the beginning'', which we fit under a new code ``Concerns over AI-generated content and fairness in model behavior'': (IC119). 
For fairness promotion techniques (RQ4), saturation was observed at participant P12, who shared an XAI-related fairness techniques for model interpretability to promote fairness during AI lifecycle, where all the quotes from next participants P13--P26 were aligned with the code \emph{Absence of formal techniques or protocols (IC123)}, conveying a pattern that all the remaining participants do not rely on techniques (e.g., ``No techniques or protocols'': P4; ``...just through data observation, no techniques'': P13; ``No techniques...just data observation'': P26), initially reported by P4 for which the code was initially generated and further shared by the rest.}

Consistent with prior empirical studies where data saturation was reported within the first 12 interviews \cite{guest2006many}, and systematic reviews that have documented saturation ranges from approximately 9 to 17 interviews, with an average near 13 interviews \cite{guest2006many, fusch2015we, mwita2022factors}, we observed that for RQ1 and RQ4 thematic codes stabilized early. \imp{However, for the remaining RQs (RQ2 and RQ3), new nuances continued to emerge later, reaching P24 for RQ2 and P26 for RQ3, reflecting participant heterogeneity and multiple thematic codes.}

\raggedbottom

\subsection{Data Analysis}\label{TA}
In this section, we describe the data analysis technique we applied, illustrating each step with an illustrative example from a specific aspect of RQ1 (definition of fairness in AI development). We demonstrate how the technique was systematically followed for each research question, by answering the corresponding sub-questions (see Section~\ref{rqs}), as demonstrated in the interview guide design in Section~\ref{guide}.

Recall, we used a Thematic Analysis~\cite{clarke2017thematic} to systematically identify common patterns (a.k.a., \textit{themes}) across our \total participants. 
\imp{For each RQ, for each associated interview sub-question (see Appendix~\ref{appendix}), the second author first selected representative quotes reflecting distinct shared experiences by our participants and organized them using a \textit{card sorting} approach\cite{spencer2009card} to generate initial codes. The first author then went through all quotes from the interviews and validated the final set of representative quotes finalized by the second author, and independently reviewed and refined the associated codes, identifying any missing concepts. The first author afterwards developed candidate, refined, and final themes by grouping related codes, consulting the third author for expertise, and holding consensus meetings with the second author at each stage to review, discuss, and resolve any disagreements. Consistent with reflexive TA and its adoption in the literature~\cite{smith2025pragmatic,ryan2023integrating}, we did not compute inter-rater reliability metrics, as rigor was ensured through iterative cross-checking and analytic discussion rather than parallel independent coding. All authors have experience in AI fairness research, including a prior gray literature investigation, and the first author has prior experience conducting interviews with industry practitioners. Additionally, the third author, with expertise in qualitative and quantitative empirical studies, contributed to validating the interview guide and advising on the study design. This combined expertise supported rigorous coding, thematic development, and analytic rigor.}

We followed Braun and Clarke’s six-phase TA framework~\cite{clarke2017thematic}, valued for its flexibility and ability to capture nuanced participant experiences, as follows:
\begin{itemize}
    \item \textbf{Data Familiarization} through repeated reading of transcripts and taking notes. Our collected transcripts are verbatim and the records are anonymized. The \emph{initial notes} we took helped capture the first insights and the recurring observations. For example, after multiple readings of the transcripts, we developed a preliminary understanding that participants focus more on delivering a tool with effective ML / AI models in terms of high accuracy (mostly through \emph{F1-score}), meeting client needs, and while they are somewhat knowledgeable about fairness, they often confuse it with model effectiveness.
    \item \textbf{Initial code Generation} by systematically identifying \emph{quotes} (meaningful data segments), and assigning them to precise, descriptive \emph{initial codes}.
    \begin{table}[!h]
\centering
\caption{Fairness definitions in AI/ML with representative participant quotes and codes}
\renewcommand{\arraystretch}{1}
\begin{tabular}{p{9cm}|p{4cm}}
\hline
\rowcolor{gray!30} 
\textbf{Quote} & \textbf{Initial Code} \\
\hline
``...fairness in this context is about... responsible AI products... ensure we are not treating people with different accuracy... gender, age... we talk about protected attributes.'' \textbf{(P2)} & Equal treatment across protected attributes \textbf{(IC1)} \\
\hline
\imp{``Data used to train the model should be balanced towards all classes (should capture all classes)''} \textbf{(P17)} \newline ``First thing...is the fairness of data because data is the backbone of the project...we should balance the amount of data...to meet the scope of the project''\textbf{ (P3)} & Data fairness through balance \textbf{(IC2)} \\
\hline
``...fairness is more of an ethical term... I don't have a quantification for it. ... how to be fair from development to the production and how to report everything ethically.'' \textbf{(P4)} & Ethical responsibility across development and production \textbf{(IC3)} \\
\hline
``Fairness is the quality of making judgments that are free from discrimination.'' \textbf{(P6)} & Unbiased, nondiscriminatory judgment \textbf{(IC4)} \\
\hline
``Technically fairness could be like key setting... same settings for all models...how the model is transparent in terms of 
diving deep into the commonly used 'black box'... dive deeper to get key factors that influence the model'' \textbf{(P9)} & Consistent model settings and transparency \textbf{(IC5)} \\
\hline
``...the ability of the ML model to generalize decisions... ensuring the model performs equally well across all regions regardless of data quality difference, meteorological conditions or region accessibility...'' \textbf{(P13)} & Model generalizability across diverse conditions \textbf{(IC6)} \\
\hline
``It means data fairness to me... using fair data and unbiased data to feed the model.'' \textbf{(P15)} & Use of fair and unbiased data \textbf{(IC7)} \\
\hline
``there should not be any bias from feature perspective while including all features... also when removing one or many of sensitive features, still the model shall remain fair.'' \textbf{(P19)} & Feature-level fairness and robustness against bias from sensitive features \textbf{(IC8)} \\
\hline
\imp{``how your model is performing on different kinds of data... there is equality between genders, classes.... It does not have to be abstract but it has to predict something on broad level'' \textbf{(P10)}}
``...can't explain fairness alone...need to include bias in the def... fairness is treating all categories of data equally.'' \textbf{(P21)} & Equal treatment of all data categories \textbf{(IC9)} \\
\hline
\imp{``coverage parity...model needs to be fair towards all classes, minority classes (e.g., women, specific ranges of age) '' \textbf{(P11)}} \newline
``Fairness is when data is accurate... if data is fair, the model is fair in the pipeline... fairness in terms of expected prediction or classification outcome... It's all about accuracy.'' \textbf{(P23)} & Fairness as accuracy of data and related model predictions \textbf{(IC10)} \\
\hline
``Any model or AI product that's developed has no bias... make sure that this has no bias related to sex or special abilities or anything like that.'' \textbf{(P25)} & Absence of bias towards protected groups \textbf{(IC11)} \\
\hline
\end{tabular}
\label{fairness_def}
\end{table}

    For instance, Table~\ref{fairness_def} illustrates some representative quotes from our participants. The collected quotes have been used to generate initial codes, w.r.t., fairness definition in AI/ML models. Quotes include ``...fairness in this context is about... responsible AI products... ensure we are not treating people with different accuracy... gender, age... we talk about protected attributes.'' (P2), ``First thing...is the fairness of data because data is the backbone of the project...we should balance the amount of data...to meet the scope of the project'' (P3), and ``...fairness is more of an ethical term... I don't have a quantification for it. ... how to be fair from development to the production and how to report everything ethically.'' (P4). These quotes reflect fairness as equal data treatment, data balance, and ethical responsibility (see initial codes \textbf{IC1}, \textbf{IC2}, and \textbf{IC3} under the `Initial Code' column), offering core practitioner views on the concept.
    \item \textbf{Theme search} by grouping the initial codes into broader thematic categories, known as \emph{candidate themes}.
    For instance, based on the initial codes generated in the previous step, we identified broader patterns and grouped these initial codes into four candidate themes, as depicted in Table~\ref{tab:theme_search}. 
       \begin{table}[H]
        \centering
        \caption{Candidate Themes and the Corresponding Initial Codes}
        \renewcommand{\arraystretch}{1.2}
        \begin{tabular}{p{3.7cm} | p{9.3cm}}
        \hline
        \rowcolor{gray!30}
        \textbf{Theme} & \textbf{Includes Initial Codes }(IC) \\
        \hline
        Data Fairness & Data fairness through balance \textbf{(IC2)}, Use of fair and unbiased data \textbf{(IC7)} \\
        \hline
        Model Fairness & Equal treatment across protected attributes \textbf{(IC1)}, Model generalizability across diverse conditions \textbf{(IC6)}, Absence of bias towards protected groups \textbf{(IC11)}, Model robustness to sensitive attribute influence in decisions \textbf{(IC9)}, Feature-level fairness and robustness against bias from sensitive features \textbf{(IC8)} \\
        \hline
        \text{Data and Model} \text{Fairness} & Fairness as accuracy of data and related model predictions \textbf{(IC10)} \\
        \hline
        \text{Ethical and Operational} Fairness & Ethical responsibility across development and production \textbf{(IC3)}, Unbiased, nondiscriminatory judgment \textbf{(IC4)}, Consistent model settings and transparency \textbf{(IC5)} \\
        \hline
        \end{tabular}
        \label{tab:theme_search}
        \end{table}   
    \item \textbf{Theme review} to adjust the candidate themes (possibly by refining, discarding, or adding) and validate coherence and consistency, resulting in \emph{refined themes}. 
    We revisited our candidate themes w.r.t., the full transcripts we collected from our participants. 
    For example, we found an overlap between i) Data Fairness and ii) Data and Model Fairness because they both involve how the quality and fairness of the input data relate directly to the model's outcomes, so we merged both candidate themes into \emph{Data and Outcome Fairness} refined theme. Model fairness and ethical and operational fairness candidate theme, however, are kept separate due to their distinct focus on the model behavior and ethical responsibilities. This process created clearer, more coherent themes reflecting fairness in AI/ML.
    
    \item \textbf{Theme definition and Naming} to capture their essence, leading to \emph{final themes}.
    After revision of the refined themes, we renamed the Data and Outcome Fairness theme to \emph{Data Quality and Outcome Fairness}, reflecting the importance of using balanced, unbiased, and representative data, and ensuring fair and accurate model results for all groups. To do so, this theme captures data-related concerns such as dataset diversity, data completeness, and their direct impact on the model predictions.
    We retained the \emph{Model Fairness} and \emph{Ethical and Operational Fairness} themes with clarified focus.
    While the former theme mainly concerns the model’s behavior after the data was fed to the model, focusing on equitable treatment, bias mitigation, generalizability, and transparency, the latter covers responsible practices throughout AI development and deployment, emphasizing nondiscrimination, accountability, and adherence to fairness policies.
\begin{table}[H]
\centering
\caption{Final themes for answering RQ1.}
\begin{tabular}{p{6.5cm} p{4.5cm} p{1.5cm}}
\toprule
\rowcolor{gray!40}
\multicolumn{3}{c}{\textbf{Themes for RQ1}} \\
\midrule
\multicolumn{3}{c}{\textbf{Definition of fairness in AI models}} \\
\rowcolor{gray!25} \hline
\textbf{Final Theme} & \textbf{\imp{Representative.P}} & \textbf{\# \imp{Total.P}} \\
\midrule
Data Quality and Outcome Fairness & P3, P15, P17, P18, P23 & 12/26 \\ 
Model Fairness & P1, P2, P10, P13, P19, P21, P25, P26 & 10/26  \\ 
Ethical and Operational Fairness & P2, P4, P6, P9 & 4/26 \\ 
\midrule
\multicolumn{3}{c}{\textbf{Examples of fairness concerns in AI models}} \\ \hline
\rowcolor{gray!25}
\textbf{Final Theme} & \textbf{\imp{Representative.P}} & \textbf{\# \imp{Total.P}} \\
\midrule
Data Quality, Balance, and Representation & P1, P5, P6, P14, P15, P26 & 13/26 \\
Impact of Sensitive Attributes on Model Fairness & P9, P19, P20 & 4/26 \\
Model Fairness Testing and Assessment & P11, P12, P21, P23 & 3/26 \\
\bottomrule
\end{tabular}
\label{rq1themes}
\end{table}
    
    \item \textbf{Report production} by weaving themes and quotes into a coherent narrative.
    The final report was carefully crafted in the last stage of the TA. In this phase, we systematically developed and presented the findings by directly addressing research questions RQ1--RQ4, as detailed in Sections~\ref{rq1}--~\ref{rq4} . The final report weaves together the identified themes into a coherent narrative, supported by authentic participant quotes that genuinely reflect their perspectives and experiences in my context.
\end{itemize}
We document the key steps of the TA in an Excel sheet~\cite{methodology_Excelsheet} to improve clarity and traceability of emerging patterns between participants and research questions.
\section{Results}\label{eval}
In this section, we report the findings of our four research questions (see Section~\ref{rqs}), providing analysis, insights, and key conclusions drawn from our qualitative analysis of fairness requirements in AI/ML-based software. We articulate the themes that we identified from the interview
responses and present the findings for each theme, supported by direct quotations from the responses of the participants. 
\imp{For each RQ, we summarize the thematic analysis results in Tables~\ref{rq1themes}, \ref{rq2themes}, \ref{rq3themes}, and \ref{rq4themes} using three columns. Column `Final Theme' presents the themes derived from participants’ responses, and Column `Representative.P' lists participants whose quotes are shown as illustrative examples. Column `Total.P' indicates the total number of participants supporting each theme. Some quotes are not shared to protect privacy and respect participants’ choice not to share full responses. \\
For some participants, the answer to a single sub-question includes more than one quote, with each quote coded, then the latter being assigned to a single theme (see details about the TA in Section~\ref{TA}) corresponding to the dimension it addresses (see P2 quotes in defining fairness in AI/ML context in Section~\ref{def}).
Conversely, some responses did not directly answer the question or indicated that no practice currently exists, and thus were not coded and assigned to any theme. For these reasons, the number of participants reported for the total themes of each RQ may not always sum to exactly the total of \total participants, exceeding or falling below that total when a participant contributes to multiple themes or not all quotes address the question.}

\minor{Additionally, for each RQ, we complement the TA findings with a quantitative summary box (see the final summaries at the end of Sections~\ref{rq1} -- \ref{rq4}) that reports the frequency of participants supporting each theme. For each theme under each sub-question, we traced back the specific participants whose shared quotes fit under that theme and cross-referenced their roles, seniority levels, and regions in which projects were developed from Table~\ref{demographics}. This allowed us to observe whether a given theme was confined to a particular subgroup (e.g., only junior developers, or only participants from a specific region) or whether it was supported by participants spanning diverse roles, seniority levels, and/or regions (e.g., theme adopted by both junior and senior practitioners across four continents), thereby assessing the breadth of each theme across the sample. We refer to this procedure as cross-cut comparison.}

\subsection{RQ1: How aware are AI/ML professionals of key concepts of fairness in AI/ML software development, and how fairness has been reflected in their work?}\label{rq1}
As depicted in Table~\ref{rq1themes}, we analyze the understanding of fairness by software professionals in the AI/ML context, providing concrete examples of fairness concerns encountered during the software development process in projects assigned to them, based on their experience.
\imp{Following the TA process~\cite{clarke2017thematic} resulted in six final themes extracted from the different sub-questions of RQ1 (see Table~\ref{rq1themes}).}

\subsubsection{Definition of fairness in AI / ML context}\label{def}
\imp{Our participants demonstrated varying conceptualizations of fairness in AI/ML systems, which generally evolved during the discussion from an initially vague understanding of the term toward more nuanced interpretations.}

At the beginning of the interviews, most participants demonstrated an implicit understanding and awareness of fairness in AI/ML contexts; despite initially conflating fairness with model accuracy and robustness, by the end of the discussion, their responses reflected a more nuanced understanding across multiple fairness dimensions. For instance, while P20 initially said ``first time I hear about fairness'', after further discussion and guidance, he demonstrated an understanding of fairness without necessarily knowing the term, stating that it is the ``ability of the model to fairly learn from historical data'' and that ``fairness is trying to make the model robust, without having sensitive attributes intervene in the decisions''.

More in detail, most participants define fairness in AI / ML context mainly through three lenses: \emph{data quality and outcome fairness}, \emph{model fairness}, and \emph{ethical or operational fairness,} while some describe fairness in ways that encompass multiple perspectives simultaneously. 

\paragraph{\textbf{Data Quality and Outcome Fairness}}

Some participants \imp{define fairness in terms of} \emph{data quality and outcome fairness}, focusing on the need of using balanced, accurate, and unbiased data to ensure that AI/ ML models perform fairly across all groups (e.g., P3, P15, P17, P18, P23).
For example, P3 said:  \\
\begin{myquote}
    {
    ``First thing when thinking about fairness is the fairness of  data because data is the backbone of the project...
we should balance the amount of data that we could utilize to meet the scope of the project''
    }
\end{myquote}
\imp{This view suggests that several participants associate fairness primarily with the representativeness and balance of datasets, assuming that equitable outcomes largely depend on minimizing bias during data collection and preparation.}

\paragraph{\textbf{Model Fairness}}

Others relate fairness to \emph{Model Fairness}, with a focus on avoiding bias related to sensitive attributes and ensuring equitable model behavior without discrimination (e.g., P1, P2, P10, P13, P19, P21, 25, P26).
For example, P19 claimed that fairness is strictly associated with the model behavior, regardless of the data fed to it, as follows:  \\
\begin{myquote}
    {
     ``There should not be any bias from feature perspective, while including all features... also when removing one or many of 
sensitive features, still the model shall remain fair.''
    }
\end{myquote}

\imp{This perspective highlights a model-centric understanding of fairness, where participants emphasize the overall model behavior and the impact of input features on it, rather than data composition.}

\paragraph{\textbf{Ethical and Operational Fairness}}
Fairness is also defined as \emph{Ethical and Operational Fairness} for others, where fairness is primarily highlighted as an ethical responsibility that requires transparency and accountability throughout the development and deployment of AI (e.g., P2, P4, P6, P9).
For instance, P4 said:  \\
\begin{myquote}
    {
     ``Fairness is more of an ethical term, so I don't have a quantification for it...how to be fair from development to the production and how to report everything ethically.''
    }
\end{myquote}

\imp{This framing suggests that some participants perceive fairness not only as a technical property of data or models, but also as a broader ethical and organizational practice requiring responsible development, transparency, and accountability.}

\imp{Recall that a single participant’s quotes from the answer to the same question can be coded then fit under more than one theme. For example, in defining fairness in AI models, P2 stated ``look at fine metrics to ensure that we are not treating people with different accuracy...concerning many aspects like gender, age… we talk about the protected attributes...'', which fits under \emph{Model Fairness} theme, while the same participant's response also includes ``fairness in this context is about...responsible AI products...'', reflecting \emph{Ethical and Operational Fairness} theme.}

\subsubsection{Examples of Fairness Concerns in Software Development Process}

\imp{Our participants reported several concrete situations from AI software development where fairness concerns emerge in practice. These examples highlight how fairness challenges arise across different stages of the AI pipeline, particularly in relation to \emph{data quality, balance, and representation}, the \emph{impact of sensitive attributes on model fairness}, and \emph{model fairness testing and assessment} practices.}

\paragraph{\textbf{data quality, balance, and representation}}

The participants shared a variety of real-world examples that highlight concerns about fairness during software and AI development. Many discussed data-related issues such as dataset imbalance (e.g., P1, P6), lack of representativeness (e.g., P5, P14, P15) and challenges arising from wrong, synthesized, or AI-generated data (P6, P26). 
For instance, P5 said:  \\
\begin{myquote}
    {
     ``The dataset from the U.S. doesn't match with the dataset for black people dataset in Ecuador... difficult to recognize the features of black people. ... people with dark skin weren't  recognized well... the model always makes mistakes for black people''
    }
\end{myquote}

\imp{This illustrates a fairness concern in practice, where the AI model performed poorly for underrepresented groups in the dataset.}

\imp{In terms of data imbalance-related fairness concerns, P6} added:  \\
\begin{myquote}
    {
    ``... \imp{...majority of people have high income… a few have low income... difficult to train the model... participants at all levels have to be equal...}''
    }
\end{myquote}

\imp{This example highlights a fairness concern in practice, where income imbalance in the dataset created difficulties in training the model fairly across all participant groups.}

\imp{Through such datasets that fail to capture sufficient diversity across demographic groups, the identified fairness concerns indicate an awareness that biased or incomplete data distributions can directly translate into unequal model performance across populations.}

\paragraph{\textbf{Impact of Sensitive Attributes on Model Fairness}}

A few participants emphasized the impact of sensitive attributes on model fairness in practice (e.g., P9, P19, P20), including the removal / inclusion of features like gender and race, affecting fairness.
For instance, P19 said:  \\
\begin{myquote}
{``...we removed sensitive attributes like city and gender and found the model was more fair without them...''
}
\end{myquote}

\imp{This shows how sensitive attributes can impact the model fairness and how participants took concrete steps to mitigate bias in practice.}.

\paragraph{\textbf{Model Fairness Testing and Assessment}}

Finally, a few participants (P11, P12, P21) described some model fairness testing and assessment practices, such as using Explainable AI (XAI) techniques to identify bias sources (P12), measuring parity metrics (P11), and performing balanced testing between different groups to evaluate model performance (P21). 
For example, P21 said: \\
\begin{myquote}
    {
    ``...same query  was given to Chatbot... test cases from Christians  and Muslim... when results are equal, then we say Chatbot is fair'' (P21)
    }
\end{myquote}

\imp{This example illustrates how participants evaluated model fairness in practice by comparing outcomes across demographic groups to ensure equitable treatment.}

\imp{Overall, these examples collectively demonstrate not only participants' awareness of AI/ML fairness concepts, but also how fairness considerations were reflected in their reported development practices across different stages of the AI pipeline, such as identifying dataset imbalances, ensuring representativeness, analyzing or removing sensitive attributes, and evaluating subgroup performance.}

However, we observed that while most participants shared insightful examples aligned with fairness concepts, a minority expressed a view that equates fairness simply with overall model accuracy, \imp{showing a misunderstanding regarding the distinction between AI/ML fairness metrics and conventional model accuracy indicators} as illustrated by P23's statement:  \\
\begin{myquote}
    {
    ``
...in covid-19 dataset... if accuracy is greater than 80\%, then the model is fair''.
    }
\end{myquote}
Further, P11 demonstrated knowledge of fairness-related metrics like computing the False Positive (FP) gaps across groups, but also showed some confusion by including the \emph{F1-score} as a main fairness metric in another context.

\begin{summarybox}[RQ1 Summary]
Our participants showed an implicit understanding of fairness in AI/ML software, progressing from initial confusion of fairness with accuracy and robustness to recognizing its multiple dimensions: data quality and representation, model behavior w.r.t., sensitive attributes, and ethical accountability. Real-world concerns included dataset imbalance, under-representation of minority classes/ groups, and sensitive attribute challenges. While most participants aligned with established fairness concepts, a minority confused fairness with the model accuracy or showed fairness metric ambiguity, indicating a need for further training to clarify these distinctions.
\end{summarybox}

\paragraph{\textbf{\imp{Quantitative Summary of RQ1 Observations Across Participants}}}

\imp{As shown in Table~\ref{rq1themes}, the majority of participants (12/26) defined AI/ML fairness in terms of data quality, including early- to mid-career AI/ML researchers, AI/ML engineers, researchers, and consultants, along with a software project manager from countries spanning Asia, Africa, North America, and Europe.
Fewer participants (10/26) defined AI/ML fairness in terms of model behavior, including early-to senior-level AI /ML researchers / engineers / data scientists and early-career R\&D specialists, from countries across North and South America,  Asia, and Africa.
Only four participants defined AI fairness as ethical and operational, representing two data scientists ((one mid-level, one senior)), one ML software developer, and a computer vision engineer (both mid-level) across North America, South America, and Europe.}

\imp{Regarding shared examples of fairness concerns, the majority of participants (13/26) at early to senior-level shared issues related to data quality, balance, and representation, including AI/ML researchers, engineers, and consultants spanning early- to senior-level experience across North America, South America, Asia, Europe, and Africa.
Fewer mid-level participants (4/26) mentioned sensitive attribute-related concerns, including AI/ML researchers, engineers, and R\&D specialists spanning South America, North America, Asia, and Europe.
Finally, four participants at early to senior-level shared fairness testing and assessment-related issues, including AI/ML researchers and engineers from Africa and Asia.
}

\begin{summarybox2}[Quantitative Summary of RQ1]
\imp{Quantitative analysis of RQ1 findings shows that definitions and concerns of AI/ML fairness are predominantly data-centric, followed by model-focused and ethical/operational perspectives. These patterns are observed across all seniority levels from early- to senior-career participants, across diverse roles including AI/ML researchers, engineers, data scientists, R\&D specialists, and consultants, and across multiple continents, including North America, South America, Europe, Asia, and Africa. This indicates that both the perception and articulation of fairness, as well as the concerns raised, are broadly shared and not restricted to specific career stages, roles, or regions.}
\end{summarybox2}

\subsection{RQ2: How are fairness concerns translated into fairness requirements? And how are the latter refined, documented,
and interpreted by stakeholders in AI/ML software development process?} \label{rq2}

For this research question, we explore how fairness concerns are transformed into actionable fairness requirements, the processes used to collect and handle them in the AI lifecycle, how they are documented, and the roles of different stakeholders in interpreting these fairness requirements in AI software development.
Answering RQ2 resulted in 16 final themes extracted from the different sub-questions in Table~\ref{rq2themes}.
In the following, we report on the evaluation of each of the sub-questions of RQ2: 

\subsubsection{Translation of Fairness Concerns into Requirements}

\imp{While some participants did not provide responses that clearly explain how fairness concerns are translated into requirements, most of our participants described how fairness concerns are translated into actionable requirements through a process combining the \emph{quantification, formalization, and documentation of fairness concerns} within development artifacts, \emph{data-driven fairness assessment and mitigation}, and \emph{stakeholder- and expert-informed fairness definition.}}

\paragraph{\textbf{Quantification, Formalization, and Documentation of Fairness Concerns}}

\imp{A key aspect of translating fairness concerns into actionable requirements involves formalizing and documenting them within development artifacts to make fairness goals measurable and enforceable (e.g., P1, P11, P21).
For instance, formalization tools like checklists, model cards, and documentation in JIRA help track and operationalize these requirements throughout the SDLC, as stated by P1: \\
\begin{myquote}
 {
``...sprint planning templates ensuring fairness were discussed during backlog grooming and sprint retrospectives... we designed model cards to address demographic skew and evaluate results broken down by gender, age, and region... we documented everything in JiRA''
}
\end{myquote}
}
\imp{This demonstrates that some participants actively use structured documentation and formal tools to embed fairness into the AI development lifecycle, ensuring that abstract fairness objectives are systematically captured, tracked, and operationalized. This suggests that without such scaffolding, fairness goals risk remaining informal and unenforceable across AI development stages.}

\paragraph{\textbf{Data-Driven Fairness Assessment and Mitigation}}

Many participants (e.g., P3, P4, P6, P10, P15, P18, P19, P22--P24) described a process in which observed data-related fairness concerns such as data imbalance, demographic skew, or sensitive attributes are explicitly translated into actionable requirements. This process often begins with data-driven assessment, where data imbalance or bias is identified, documented, and quantified using defined metrics.

For instance, P6 quoted: \\
\begin{myquote}
 {
``Data balance impacts directly on the fairness of the models...''
}
\end{myquote}

Afterwards, participants start to reflect on how to mitigate such fairness concerns, where P10 said: \\ 
\begin{myquote}
 {
``...start to reflect on techniques to mitigate imbalance, such as mathematical operations and sampling for oversampled data, using open source libraries like those from IBM and Microsoft''.
}
\end{myquote}
Based on these observations, requirements are therefore defined, for example as constraints on model predictions \imp{(``...first define fairness...translated these concerns into proper sentences: the model shall consistently predict the traffic level across all road types'': P22)} or selected features that are fed to the proper AI models \imp{(``observing data like gender/city...fairness requires a model...not biased towards these features'': P19)}.

\imp{These examples indicate that participants approach fairness as an iterative, data-informed process: first identifying potential biases, then designing mitigation strategies, and finally translating these insights into concrete, actionable requirements. This suggests that fairness requirements are not defined upfront but emerge iteratively from the data itself, making data quality and availability a prerequisite for meaningful fairness engineering}

\paragraph{\textbf{Stakeholder- and Expert-Informed Fairness Definition}}

Stakeholder and expert input further informs the mapping from fairness concerns to requirements (e.g., P17, P25), ensuring domain knowledge, ethical considerations, and regulatory guidance. For instance, p25 quoted the following: \\

\begin{myquote}
{
``...meet with the finance department to define requirements ensuring no bias affects the model...for example, excluding special abilities as a feature.''
}
\end{myquote}

\imp{This demonstrates that fairness is not only a technical problem but also a socio-technical one, where expert judgment and stakeholder perspectives are critical to defining meaningful and contextually appropriate requirements.}

Overall, results indicate that fairness concerns are converted into requirements through an iterative process, combining data analysis, stakeholder input, and formal documentation, making abstract fairness goals measurable and enforceable in practice.

\subsubsection{Collection of Fairness Requirements}

\imp{Participants described the collection of fairness requirements through a combination of stakeholder and expert involvement, metric-driven elicitation, and data management practices, highlighting how requirements are defined and informed in practice.}

\paragraph{\textbf{Stakeholder and Expert Involvement in Fairness
Requirements}}

\imp{A central mechanism for collecting fairness requirements is engaging stakeholders and domain experts to surface concerns and guide requirement definition.}
 
\imp{More in detail,} teams employed a combination of different stakeholder consultations (e.g., P1, P10, P11, P14, P15, P19, P20--P22),
\imp{from a domain-expertise perspective through regular meetings (e.g., ``weekly meetings with supervisors...during discussions, we get instructions and requirements from superiors'', quoted P11)}, and from data-expertise perspective (e.g., P14, P21), highlighting ongoing discussions about various aspects, such as sensitive attributes, where, for instance, P21 stated the following: \\
\begin{myquote}
{
``Once we get data, we discuss with stakeholders... they share their concerns about sensitive attributes... we write the fairness requirements to check if they are satisfied on the chatbot.''
}
\end{myquote}

\imp{Further, stakeholder engagement did not remain restricted to experts but also involved all team members (``...encourage all staff to share their opinions...brainstorming sessions among team members'', quoted P10) and clients (``based on discussions with the client, we collect those requirements'', added P15).}
 
\imp{These examples show that input from different stakeholders is critical for surfacing contextual fairness concerns, aligning requirements with domain knowledge, and ensuring practical relevance.}

 \paragraph{\textbf{Metric-Driven Fairness Requirement Elicitation}}
 
 \imp{Metrics operationalize fairness concerns by making abstract issues like skewed distributions measurable, which in turn surfaces concrete, actionable requirements. This is further supported by P1, who shared the following:} \\
 
\begin{myquote}
{
``...integrated fairness metrics into our evaluation frameworks...identify imbalances techniques to address like skewed distribution''
}
\end{myquote}
  
 \imp{This indicates that quantitative measures play a complementary role in collecting fairness requirements, helping to operationalize abstract fairness goals into measurable system properties.}

 \paragraph{\textbf{Data Management to Support Fairness}}

\imp{Data management practices further support fairness requirement collection by identifying potential sources of bias and guiding mitigation strategies before requirements are finalized.} 

\imp{More in detail,} various data management-related tasks such as data visualization to detect bias (e.g., P3), data distribution inspection to identify class imbalance (e.g., P1), data cleaning (P4), outlier removal (P6), undersampling or augmentation to balance classes (P8, P12), and selective inclusion of attributes (P23, P25) to mitigate potential unfairness (e.g., ``select attributes that might cause unfairness then...decide on whether...to include them'', quoted P23), contributed to the full set of collected fairness requirements in practice.

\imp{These practices demonstrate that fairness requirement collection is closely tied to proactive data handling, ensuring that technical data-level interventions align with the fairness goals surfaced by stakeholders and metrics.}

\imp{Overall, fairness requirement collection is a multi-faceted process combining stakeholder input, quantitative metrics, and proactive data management to translate abstract fairness goals into concrete, actionable requirements.}

\subsubsection{Handling Fairness Requirements in Software Development Process}
\imp{Unlike collection, handling fairness requirements involves sustaining and enforcing them as the AI system evolves.}
Most of the participants shared how fairness requirements are handled from three different perspectives: \emph{stakeholder roles and oversight in fairness}, \emph{iterative fairness requirement refinement}, and \emph{fairness-oriented data and model management, reflecting a multi-faceted approach to managing and enforcing fairness requirements throughout the AI development process} (see Table~\ref{rq2themes}). 

\paragraph{\textbf{Stakeholder Roles and Oversight in Fairness}}

\imp{Stakeholder involvement and supervisory oversight play a critical role in ensuring that fairness requirements are properly defined, monitored, and aligned with organizational objectives (P1, P10).}

For instance, P10 shared a supervisor oversight stating that ``...supervisor looks into fairness concerns, and assigns tasks to us''.

\imp{This demonstrates that fairness handling relies on expert judgment and hierarchical oversight, where senior roles guide fairness requirement interpretation, task delegation, and accountability throughout model development.}

\paragraph{\textbf{Iterative Fairness Requirement Refinement}}

\imp{Our participants described an iterative process of continuous refinement as AI systems evolve and new fairness concerns emerge.}
\imp{More in detail,} fairness requirements were iteratively refined (e.g., P15, P18, P20--P22, P24, P26) based on observed data issues and model behavior, with repeated testing, feedback, and prompt updates for LLMs (P21, P24). 

For instance, P21 commented on the use of LLMs: \\

\begin{myquote}
 {
``...launch fairness queries...we rephrase many times as chatbot shows different behaviors...
 use many different syntax while preserving the semantics... 
we pick the queries from which chatbot behaved less sensitively towards sensitive attributes''.
}
\end{myquote}

\imp{This highlights that requirement handling is a dynamic process, where iterative testing and adjustments ensure that fairness considerations remain responsive to actual model outputs.}

\paragraph{\textbf{Fairness-Oriented Data and Model Management}}

\imp{Participants also integrated data- and model-centric strategies to support fairness, ensuring that technical interventions align with fairness requirement objectives.}

\imp{More specifically,} many data-centric strategies (e.g., P3--P6, P13--P15, P17, P19, P23, P25) were considered as part of the fairness requirement handling in software development such as data cleaning, annotation, sanitization, balancing, and domain-specific adaptation (e.g., ``...using data augmentation...delete outliers...'', quoted P6), \imp{along with model effectiveness-based refinement (e.g., ``...data analysis... consultants define requirements based on their expertise...refining requirements based on model effectiveness...'': P15).}
\imp{These practices indicate that fairness handling encompasses concrete technical steps on the data and model side, complementing oversight and iterative refinement to create a holistic approach to promoting fairness in AI systems.
}

Together, these practices illustrate an integrated approach that combines expert oversight, iterative fairness requirement refinement, and fairness-oriented data and model management to promote fairness in AI models, \imp{going beyond isolated technical fixes to sustain fairness throughout the development process.}.

\subsubsection{Documentation of Fairness Requirements}
\imp{Participants described documenting fairness requirements through \emph{shared digital platforms}, \emph{fairness artifacts}, \emph{data and execution records}, but also reported \emph{lack of official documentation} in many cases, reflecting a mix of formal and informal practices in capturing fairness knowledge.}

\paragraph{\textbf{Shared Digital Platforms}}

\imp{A common approach to documenting fairness requirements involves using shared digital platforms to store, organize, and share relevant information.
To provide a concise overview of these documentation practices, Table~\ref{tools_doc} illustrates the different documentation tools / approaches that fit under each documentation theme, along with the total number of participants supporting each fairness requirement documentation theme from Table~\ref{rq2themes}. As some participants reported multiple fairness requirement documentation practices or tools, the summed total across all documentation themes exceeds the number of participants.}

\begin{table}[h!]
\centering
\caption{\imp{Summary of Fairness Requirements Documentation Methods Across Participants}}
\label{tools_doc}
\begin{tabular}{p{5cm}|p{8cm}}
\hline
\textbf{\imp{Theme (\#P)}} & \textbf{\imp{Representative Tools/Platforms}} \\
\hline
Shared Digital Platforms (11) & Jira, Google Drive, private repositories, GitHub, OneDrive, Scrum docs, internal collaborative platforms, Google Docs, Confluence \\
Fairness Artifacts (4) & Ethics/regulatory documents, dictionaries, logs and company databases, general product requirements \\
Data and Execution Records (3) & Execution records \\
Lack of Official Documentation (10) & No documentation; internal meetings \\
\hline
\end{tabular}
\end{table}

More in detail, most documentation is stored on formal platforms, including Jira (P1), Google Drive (P15, P23), private repositories (P15), GitHub (P18), OneDrive (P18), Scrum docs (P15), internal collaborative online documentation platforms such as wiki and Confluence(P5, P7), and Google Docs (P26). 
\imp{For example, P7 shared:} \\
\imp{\begin{myquote}
 {
``...we create a Confluence page where our documentation of the AI products is listed and accessible to everyone...''
 }
 \end{myquote}
}
\imp{We also noticed that some participants reported multiple fairness requirement documentation practices or tools. For instance, P1 shared that within the company he works for, the shared digital platform Jira is used along with the model cards as fairness artificats, quoting the following:  \\
\begin{myquote}
 {
``we maintain a dedicated fairness and ethics requirements document...we also designed this model cards... we were also documenting everything in JIRA.''
}
\end{myquote}
}
\imp{
This demonstrates that participants frequently leverage collaborative platforms to centralize fairness requirements, facilitating accessibility, version control, and coordination across team members.}

\paragraph{\textbf{Fairness Artifacts}}

\imp{A few participants (e.g., P2, P4, P13) create and maintain fairness-specific artifacts to codify requirements, track compliance, and provide structured references for ethical and regulatory considerations.
}

\imp{For instance, some map fairness requirements from relevant regulations, embedding them into ethics approval documents prepared by a designated regulatory officer (P2, P4), or document them in dedicated dictionaries (P13).
For example, P4 said:  \\
\begin{myquote}
 {
``...fairness is documented through the regulatory...approved by ethics approval...ethics document is prepared by the regulatory person''.
}
\end{myquote}
}

\imp{These artifacts provide a structured, traceable record of fairness considerations, linking regulatory compliance to concrete project documentation.}

\paragraph{\textbf{Data and Execution Records}}

\imp{In addition to shared platforms formal artifacts, participants also capture fairness requirements through logs and company databases and record it as data and execution records such as P21 and P22 (e.g., ``we save the log...query and execution...'', quoted P21).}

\imp{This suggests that tracking fairness requirements may support accountability and reproducibility, as participants link documented requirements with real-world system behavior.}

\paragraph{\textbf{Lack of Official Documentation}}

In spite of the different fairness requirement documentation methods shared by some of our participants, many of them (e.g., P6, P8--P10, P20, P24) say that fairness requirements are not documented within the organizations they work in (e.g., ``No documentation... all through communication through internal meetings'', quoted P24).

\imp{This gap suggests that fairness requirements may remain inconsistent and difficult to trace across projects, posing challenges for accountability and reproducibility.}

Overall, participants use a range of methods to document fairness requirements. However, many report no formal documentation, often relying on internal meetings, making the knowledge informal, oral, and unofficially followed, \imp{highlighting the need for standardized, centralized documentation to improve traceability, accountability, and reproducibility of fairness practices across projects.}

\subsubsection{Stakeholder Roles in Interpreting Fairness Requirements}
In interpreting fairness requirements, participants highlighted the participation of multiple types of stakeholder with different roles, 
\imp{involving \emph{technical and domain-expert stakeholders}, \emph{management and leadership stakeholders}, and \emph{cross-functional and operational teams}, each contributing distinct perspectives and expertise to operationalize fairness in practice (see Table~\ref{rq2themes}).}

\paragraph{\textbf{Technical and Domain-expert Stakeholders}}

\imp{Technical and domain-expert stakeholders play a central role in mapping fairness requirements to actionable implementations and ensuring that AI/ML systems adhere to fairness principles (e.g., P3, P13, P14--P17).}
\imp{Among these, participants mentioned} software developers (``...developers implemented algorithmic fairness, integrated fairness metrics into model evaluation, and applied techniques like oversampling and threshold adjustment to mitigate bias'': P1), software testers, and ``whoever is a domain expert or knowledgeable about AI/ML is involved'' (quoted P13).

\imp{This demonstrates that technical expertise is critical for embedding fairness into algorithms and models, from developers implementing algorithms and mitigation techniques to testers and broader domain experts ensuring adherence to fairness principles.}

\paragraph{\textbf{Management and Leadership Stakeholders}}

Management and leadership stakeholders also provide oversight and strategic guidance (e.g., P4, P10, P19, P20), such as senior managers, project leads, and VPs (e.g., ``Senior engineering manager, director of the team, VP'': P19, and ``project manager / senior researcher'': P20). 

\imp{This indicates that management engagement extends fairness responsibility beyond technical teams, embedding it as an organizational priority within the broader stakeholder interpretation process.}

\paragraph{\textbf{Cross-Functional and Operational Teams}}

\imp{Cross-functional and operational teams coordinate the practical execution of fairness requirements, ensuring that tasks are implemented correctly across all relevant roles (e.g., P1, P5, P7, P22, P24).}
This includes testers, designers, clients, and labeling teams, as shared by P10 (``First project manager, team lead, then developers, designers, testers... business analyst ensures requirements are properly addressed'') and emphasized by P5 (``client...government... labeling team... testing team''). 

\imp{This shows that mapping fairness requirements into practice requires coordinated effort across diverse roles, ensuring responsibilities are clearly distributed throughout the AI development process.}

Together, these layers illustrate a structured, multi-level approach where technical, managerial, and operational expertise intersect to interpret fairness.
 \begin{table}[H]
     \centering
     \caption{Final themes for answering RQ2.}\label{rq2themes}
    \begin{tabular}{p{7cm} p{4cm} p{1.5cm}}
    \rowcolor{gray!40}
    \multicolumn{3}{c}{\textbf{Themes for RQ2}} \\
    \multicolumn{3}{c}{\textbf{Translation of Fairness Concerns into Requirements}} \\ \hline
    \rowcolor{gray!25}
    \textbf{Final Theme} & \textbf{\imp{Representative.P}} & \textbf{\# \imp{Total.P}} \\
    \midrule
    Quantification, Formalization, and Documentation of Fairness Concerns & P1, P11, P21  & 3/26 \\ 
    Data-Driven Fairness Assessment and Mitigation & P3, P4, P6, P10, P15, P18, P19, P22--P24 & 10/26 \\ 
    Stakeholder- and Expert-Informed Fairness Definition & P17, P25 & 2/26\\
    \midrule
     \multicolumn{3}{c}{\textbf{Collection of Fairness Requirements}} \\ \hline
     \rowcolor{gray!25}
     \textbf{Final Theme} & \textbf{\imp{Representative.P}} & \textbf{\# \imp{Total.P}} \\ \hline
     Stakeholder and Expert Involvement in Fairness \text{Requirements} & P1, P10, P11, P14, P15, P19, P20--P22 & 10/26 \\ 
    Metric-Driven Fairness Requirement Elicitation & P1 & 1/26 \\ 
    Data Management to Support Fairness & P1, P3--P6, P8, P12, P16, P17, P25 & 11/26\\
    \midrule
    \multicolumn{3}{c}{\textbf{Handling Fairness Requirements in Software Development Process}} \\ \hline
    \rowcolor{gray!25}
    \textbf{Final Theme} & \textbf{\imp{Representative.P}} & \textbf{\# \imp{Total.P}} \\ \hline
    Stakeholder Roles and Oversight in Fairness & P1, P10 & 2/26\\ 
    Iterative Fairness Requirement Refinement & P15, P18, P20--P22, P24, P26 & 8/26\\ 
    Fairness-Oriented Data and Model Management & P3--P6, P8, P9, P14, P17, P19, P23, P25 & 14/26 \\
    \midrule
    \multicolumn{3}{c}{\textbf{Documentation of Fairness Requirements}} \\ \hline
    \rowcolor{gray!25}
    \textbf{Final Theme} & \textbf{\imp{Representative.P}} & \textbf{\# \imp{Total.P}} \\ \hline
    Shared Digital Platforms & P1, P5, P15, P18, P23, P26 & 11/26 \\ 
    Fairness Artifacts & P13 & 4/26 \\ 
    Data and Execution Records & P21, P22 & 3/26 \\ 
    Lack of Official Documentation & P6, P8--P10, P20, P24 & 10/26 \\ 
    \midrule
    \multicolumn{3}{c}{\textbf{Stakeholder Roles in Interpreting Fairness Requirements}} \\ \hline
    \rowcolor{gray!25}
    \textbf{Final Theme} & \textbf{\imp{Representative.P}} & \textbf{\# \imp{Total.P}} \\ \hline
   Technical and Domain-expert Stakeholders & P3, P13, P14--P17 & 9/26 \\ 
    Management and Leadership Stakeholders & P4, P10, P19, P20 & 4/26 \\ 
    Cross-Functional and Operational Teams & P1, P5, P7, P22, P24 & 7/26 \\
     \bottomrule
    \end{tabular}
\end{table}

\begin{summarybox}[RQ2 Summary]
Fairness requirements in AI projects emerge from an integrated but inconsistent process that combines technical, managerial, and operational perspectives with data-driven assessment and iterative refinement. Practices such as data visualization, balancing, cleaning, augmentation, and outlier removal help translate fairness concerns into measurable requirements. Nevertheless, their application varies widely across projects. Although diverse stakeholders contribute specialized expertise, some participants remain unfamiliar with fairness requirements or rely on informal, oral communication rather than standardized documentation. These knowledge, consistency and traceability gaps underscore the need for centralized formal processes to ensure accountability and reproducibility of fairness practices.
\end{summarybox}

\paragraph{\textbf{Quantitative Summary of RQ2 Observations Across Participants}}
\imp{As shown in Table~\ref{rq2themes}, several participants (10/26) shared data-driven fairness concerns translation into requirements, primarily composed of AI researchers and AI engineers, along with an ML engineer, ML software developer, data scientist, AI consultant, and software project manager, representing mainly early- to senior-level practitioners across Asia, North America, South America, Europe, and Africa.
Other participants emphasized quantification, formalization, and documentation of fairness concerns (3/26; e.g., P1, P11, P21), including a data science manager and an early-career R\&D specialist across North America, Africa, and Asia.
Finally, stakeholder- and expert-informed fairness-based definitions were noted by 2/26 participants, with a mid-career ML researcher and a mid-career data scientist across Asia and North America. 
This suggests that most practitioners derive fairness requirements empirically from data rather than through structured elicitation or formal specification.}

\imp{The most common practice in collecting fairness requirements is through data management (11/26), primarily comprising of early- to senior-level AI/ML researchers, AI engineers, data scientists, and computer vision engineers across North America, South America, Asia, Africa, and Europe.
Stakeholder and expert involvement in fairness requirement collection is reported by ten participants, including one senior data science manager and one senior AI researcher, but mostly early- to mid-level AI/ML researchers, AI consultant, ML engineer, and R\&D specialists,  across North America, Europe, Asia, and Africa.
Metric-driven fairness requirement elicitation is the least employed technique shared by one participant only, working as a senior data science manager in North America. This reveals that fairness requirement collection relies more on experiential and data-driven judgment than formal measurement techniques.}

\imp{In terms of handling fairness requirements, fairness-oriented data and model management was the most common approach, reported by 14/26 participants, involving activities such as data collection, preprocessing, balancing, and model adjustments to ensure fairness. This group mainly comprises early- to senior-level AI/ML researchers / engineers, mid- to senior-career data scientists, spanning Asia, North America, South America, and Europe. Iterative fairness requirement refinement was reported by 8/26 participants, who described repeatedly revisiting and adjusting fairness requirements through testing, evaluation, and team discussions, consisting mainly of early- to mid-career AI/ML engineers, AI researchers, and R\&D specialists across North America, Europe, Asia, and Africa. 
Only two participants reported stakeholder roles and oversight in fairness, where supervisors or stakeholders guided or assigned fairness-related tasks during development, representing a senior data science manager and an early-career AI researcher from North America and Asia.
This shows that fairness requirements in practice is largely performed by hands-on data and model work, suggesting that organizational structures or senior oversight play a limited role in shaping fairness outcomes.}

\imp{The most common fairness requirement documentation practice is through shared digital platforms (11/26), including a senior data science manager, mid-level AI engineers, early- to mid-level AI/ML researchers, AI consultant, and R\&D specialist across North America, South America, Europe, Asia, and Africa.
Next in frequency is lack of official documentation (10/26), including mid- to senior-level AI/ML researchers and engineers, and R\&D specialists across Asia, Europe, Africa, and North America.
Finally, a small number of participants shared fairness artifacts (4/26) and execution records (3/26) as alternative documentation tools, covering senior-level AI researchers with experience on projects conducted in between Africa, North America, and Asia and an early-career R\&D specialist in Asia.
The near-equal prevalence of shared platforms and informal practices suggests that access to documentation tools does not guarantee their use, revealing a systemic documentation gap in fairness requirements management.}

\imp{In interpreting fairness requirements, technical and domain-expert stakeholders were the most frequently involved, reported by 9/26 participants, including early- to senior-level AI/ML researchers and AI consultants, with one senior AI researcher, across Asia, North America, Europe, and Africa.
Cross-functional and operational teams were involved in 7/26 cases, collaborating across roles such as one senior data science manager, early- to senior-level AI engineers / researchers across North America, South America, and Asia.
Management and leadership stakeholders were the least represented, with 4/26 participants noting that supervisors, project managers, or government authorities guided or approved fairness-related decisions.
This shows that fairness requirement interpretation is predominantly driven by technical and domain experts rather than management, suggesting that fairness decisions remain largely practitioner-led with limited top-down governance.}

\begin{summarybox2}[Quantitative Summary of RQ2]
Quantitative analysis of RQ2 findings shows that data-driven and experiential approaches dominate, from translation of fairness concerns to fairness requirements to the collection, handling, documentation, and interpretation of the latter in the AI lifecycle. Patterns are consistently observed across all seniority levels, ranging from early- to highly senior practitioners, and across diverse roles, including AI/ML researchers, engineers, data scientists, and R\&D specialists. They also span multiple continents, including North America, South America, Europe, Asia, and Africa. This indicates that fairness practices, gaps, and challenges are widely shared rather than limited to specific career stages, positions, or regions, while more structured or metric-driven approaches and top-down governance remain the exception rather than the norm.    
\end{summarybox2}

\subsection{RQ3: How do fairness concerns arise and manifest early in AI/ML projects within the SDLC? And what challenges do teams face in handling them?}\label{rq3}
\begin{table}[H]
     \centering
     \caption{Final themes for answering R3.}\label{rq3themes}
    \begin{tabular}{p{5cm} p{5cm} p{1.5cm}}
    \rowcolor{gray!40}
 \multicolumn{3}{c}{\textbf{Themes for RQ3}} \\
    \multicolumn{3}{c}{\textbf{Fairness Concerns in early SDLC Stages}} \\ \hline
    \rowcolor{gray!25}
    \textbf{Final Theme} & \textbf{\imp{Representative.P}} & \textbf{\# \imp{Total.P}} \\
    \midrule
    Data Representation Bias & P1, P4, P6, P25 & 4/26\\
    Data Imbalance and Labeling Issues & P3, P5, P8, P10, P23, P24 & 8/26\\
    Data Coverage gaps & P13 & 1/26\\
    \midrule
    \multicolumn{3}{c}{\textbf{Challenges in Identifying and Applying Fairness}} \\ \hline
    \rowcolor{gray!25}
    \textbf{Final Theme} & \textbf{\imp{Representative.P}} & \textbf{\# \imp{Total.P}} \\
    \midrule
     Data-Related Fairness Challenges & P1, P3, P5, P7, P12, P14, P22, P25 & 12/26 \\ 
    Fairness Definitions and Metrics Challenges & P1, P21, P22 & 4/26 \\ 
    Model-Related Fairness Challenges & P15, P24, P26 & 3/26 \\ 
    Time Limitation & P22 & 1/26\\
     \bottomrule
    \end{tabular}
\end{table}
\imp{In addressing RQ3, we explore how fairness concerns are raised early in SDLC stages, along with challenges encountered in identifying and applying fairness in AI software, resulting in seven final themes extracted from the different sub-questions in Table~\ref{rq3themes}.}

\subsubsection{Fairness Concerns in early SDLC Stages}
\imp{Participants highlighted that fairness concerns in early stages of the software development lifecycle primarily arise from biases and gaps in the underlying data, affecting the foundation on which AI/ML models are built.}

\paragraph{\textbf{Data representation bias}}

Participants reported early-stage fairness concerns that are mainly related to data representation, quality, and coverage. 

Examples include demographic biases in datasets such as race and gender (P1, P4, P6, P25).
For instance, P6 mentioned a case of underrepresentation of a particular high financial status as compared to the majority of data, quoting the following: \\
\begin{myquote}
 {
``...a concern of fairness because I have to explain that it's impossible to model customers that have income after 10 million dollars because 
we don't have much example; we treated as outlier''
}
\end{myquote}

Similarly, P4 shared an overrepresentation issue of a specific race in the dataset as compared to other races, stating the following: \\
\begin{myquote}
 {
``...the dataset mostly comes from the Hispanic population...was biased to the Hispanic race and eyes.''
 }
\end{myquote}
\imp{These examples show that early recognition of representation bias is critical to prevent model decisions from disproportionately favoring or disadvantaging certain groups.}

\paragraph{\textbf{Data Imbalance and Labeling Issues}}

\imp{Imbalanced or mislabeled data can exacerbate unfair outcomes and reduce the reliability of model predictions.}

Examples include messy or mislabeled data (P3, P5, P23) and class imbalance (P8, P10, P24).
For instance, P10 quoted the following: \\
\begin{myquote}
 {
``we have imbalanced data...patients are more than healthy subjects''
}
\end{myquote}

\imp{This highlights that careful data preprocessing, labeling verification, and balancing strategies are essential to mitigate early-stage fairness risks and maintain model integrity.}

\paragraph{\textbf{Data Coverage Gaps}}

\imp{Gaps in dataset coverage, such as missing information from certain regions or populations, can introduce systematic blind spots in model behavior.
Examples include reporting missing data from certain regions (``Images were not available in my area...so I had issue taking data from my region...could not feed the model with data outside my region'': P13).
Addressing coverage gaps ensures that AI/ML models are trained on representative samples, reducing the likelihood of unfair treatment for underrepresented groups.}

These issues underscore the importance of proactively identifying and addressing data-related fairness concerns prior to model development.

\subsubsection{Challenges in Identifying and Applying Fairness}
While one participant did not share any fairness limitation or challenge (P19), the remaining participants (e.g., P1, P3, P5, P7, P12, P14, P15, P21,  P22, P26) faced multiple fairness challenges (see Table~\ref{rq3themes}).
\imp{Shared challenges span data, metrics, model behavior, and practical constraints, highlighting the complexity of operationalizing fairness in AI/ML projects (see Table~\ref{rq3themes}). In the following, we elaborate on each of these challenges:}

\paragraph{\textbf{Data-Related Fairness Challenges}}

Data-related fairness challenges were commonly reported (eg., P1, P3, P5, P7, P12, P14, P22, P25), including dataset imbalance (P1, P12, P21), category underrepresentation, dataset unsuitability for a particular demographic group (e.g., ``...data wasn't enough...dark color people or enough women...so we have our own team that creates and labels data'', said P5), and restricted data visibility (``data from organization like banks is not visible...40\% visibility...'': P7;  ``...mostly about restricted access to data'': P15).

\imp{These examples underscore that without adequate, representative, and accessible data, even well-intentioned fairness requirements cannot be fully realized, necessitating proactive data curation and augmentation strategies.}

\paragraph{\textbf{Fairness Definitions and Metrics Challenges}}

\imp{Ambiguities in fairness definitions and unclear or missing metrics further complicated implementation (e.g., P1, P21, P22).}
Fairness definitions and metrics were unclear (e.g., ``...lack of a clear definition of the fairness...'': P1; ``...no clear metrics...'': P21). 

\imp{This demonstrates that operationalizing fairness requires explicit definitions and measurable criteria to guide development, evaluation, and verification processes.}

\paragraph{\textbf{Model-Related Fairness Challenges}}

\imp{Model-specific issues, including technical limitations and unpredictable AI behavior, presented additional hurdles for fairness.}
Among these technical issues, we mention AI-generated-based ones as reported by some participants (e.g.,``...hallucination issue...might affect fairness'': P15; ``...LLMs always show unfairness...'': P24 ; ``Fine tuning is a big concern...'': P26). 

\imp{These observations indicate that fairness must be addressed not only at the data level but also in model design, training, and evaluation to prevent unintended biased behaviors.}

\paragraph{\textbf{Time Limitation}}
\imp{Practical constraints, such as limited time, can restrict the extent to which fairness considerations are implemented.}
P22 shared a time constraint in addressing fairness in AI models.

\imp{This highlights that operational realities, including project timelines, can impede comprehensive fairness practices, requiring prioritization and efficiency in fairness interventions.}
 
\begin{summarybox}[RQ3 Summary]
Fairness concerns arise early in the SDLC, mainly due to data biases, class imbalance, and missing or unsuitable data for certain groups. Participants shared various challenges in identifying and/or applying fairness in AI context, such as unclear fairness definitions and metrics, restricted data access, and technical model issues such as bias persistence, and AI encountered unfairness and the corresponding fine-tuning difficulties. Time constraints further limit addressing these fairness concerns. 

Overall, gaps remain in standardizing fairness definitions, the corresponding fairness evaluation metrics, data representativeness, and support of early fairness interventions.
\end{summarybox}

\paragraph{\textbf{Quantitative Summary of RQ3 Observations Across Participants}}
\imp{As depicted in Table~\ref{rq3themes}, fairness concerns in early SDLC stages were primarily data-centric. 
Data imbalance and labeling issues were reported by 8/26 participants, highlighting challenges in balancing and annotating datasets. This theme is mainly supported by early- to mid-career AI researchers and AI engineers, along with a software project manager and a research fellow, representing Asia and South America.
Data representation bias was noted by 4/26 participants, reflecting concerns about over- or under-representation of certain groups, covering early- to highly senior-level AI/ML professionals, including ML developers and data scientists across North and South America. 
Data coverage gaps, however, were mentioned by 1/26 participant (a senior AI researcher who had hands on projects across North America and Africa), showing limited availability of relevant data in some regions.
This suggests that practitioners are more attuned to technical data quality problems than to the broader societal implications of biased or incomplete data representation.
}

\imp{Data-related fairness challenges were the most common, reported by 12/26 participants (mainly early- to senior-level AI/ML researchers, AI engineers across North America, South America, Asia, and Africa), highlighting issues such as dataset imbalance, messy or missing data, and costly data preprocessing.  
Fairness definitions and metrics challenges were noted by 4/26 participants, reflecting unclear or inconsistent interpretations of fairness across teams, mainly reported by early- to senior-level AI researchers across North America and Asia.}

\imp{Further, model-related fairness challenges were reported by 3/26 participants, primarily early- to mid-level AI researchers and an AI consultant, across North America, Europe, and Asia, highlighting difficulties in ensuring equitable model behavior.
}

\imp{Finally, time limitations were noted by one senior AI researcher from Asia as a constraint on thorough fairness evaluation.}

\begin{summarybox2}[Quantitative Summary of RQ3]
Quantitative analysis of RQ3 findings shows that fairness concerns and challenges in early SDLC stages are predominantly data-centric, with data imbalance, labeling, and representation issues most frequently reported. These patterns involve participants across the full range of seniority, from early- to highly senior-level AI/ML researchers, engineers, and data scientists, and span multiple continents, including North America, South America, Asia, Europe, and Africa. Less frequently reported issues, such as fairness definitions, model-related challenges, and time limitations, follow the same trend of cross-seniority and cross-regional involvement. Overall, these findings indicate that early-stage fairness challenges are broadly shared across roles, experience levels, and regions, rather than restricted to specific positions or locations.
\end{summarybox2}

\subsection{RQ4: How do teams implement, validate, and evaluate fairness requirements during development? And how do they 
manage trade-offs with other project goals?}\label{rq4}
\begin{table}[!h]
    \centering
    \caption{Final themes for answering R4.}\label{rq4themes}
    \begin{tabular}{p{7cm} p{4cm} p{1.5cm}}
    \rowcolor{gray!40}
    \multicolumn{3}{c}{\textbf{Themes for RQ4}} \\
    \multicolumn{3}{c}{\textbf{Techniques and Approaches for Promoting Fairness}} \\ \hline
    \rowcolor{gray!25}
    \textbf{Final Theme} & \textbf{\imp{Representative.P}} & \textbf{\# \imp{Total.P}} \\
    \midrule
     Data-oriented fairness promotion & P1, P5, P11, P25 & 7/26\\ 
    Model-oriented fairness promotion & P1, P12 & 2/26 \\ 
    Ethical/social and organizational considerations & P1 & 1/26 \\ 
    Lack of implemented fairness techniques & P4, P13, P26 & 11/26 \\
    \midrule
    \multicolumn{3}{c}{\textbf{Metrics for Validating Fairness Requirements}} \\ \hline
     \rowcolor{gray!25}
    \textbf{Final Theme} & \textbf{\imp{Representative.P}} & \textbf{\# \imp{Total.P}} \\
    \midrule
    Explicit group and individual metrics & P1, P22 & 3/26 \\
    Limited / absent fairness evaluation & P4, P18, P20, P21 & 12/26 \\
    Standard performance metrics as fairness proxy & P6, P8, P9, P11, P12, P18, P22 & 11/26 \\ 
    \hline
    \multicolumn{3}{c}{\textbf{Challenges in Validating Fairness in AI Systems}} \\ \hline
 \rowcolor{gray!25}
    \textbf{Final Theme} & \textbf{\imp{Representative.P}} & \textbf{\# \imp{Total.P}} \\
    \midrule
    Conceptual ambiguity in fairness definitions & P1, P21 & 2/26 \\
    Data Quality and representation challenges & P3, P5, P10--P12, P17, P22 & 10/26 \\
    Model-related challenges & P6, P8, P24, P26 & 4/26 \\ 
    Contextual and organizational challenges & P9, P13, P15 & 4/26 \\ 
    Evaluation framework challenges  & P25 & 3/26 \\ \hline
    \multicolumn{3}{c}{\textbf{Trade-offs Between Fairness and Project Goals (Performance, Usability, Deadlines)}} \\ \hline
    \rowcolor{gray!25}
    \textbf{Final Theme} & \textbf{\imp{Representative.P}} & \textbf{\# \imp{Total.P}} \\
    \midrule
    Fairness as a core value  & P11, P15, P17 & 4/26 \\ 
    Fairness as negotiable  & P8--P10, P12, P24, P26 & 6/26 \\ 
    Fairness excluded from decision making  & P1, P3--P5, P13 & 16/26 \\ \hline
    \multicolumn{3}{c}{\textbf{Trade-offs Between Fairness and Functional Feature Prioritization}} \\  \hline
    \rowcolor{gray!25}
    \textbf{Final Theme} & \textbf{\imp{Representative.P}} & \textbf{\# \imp{Total.P}} \\
    \midrule
    Functional features prioritized over fairness  & P1, P9, P10, P13 &  16/26 \\
    Fairness Prioritized or Balanced with Functional Features  & P7, P15--P17, P24 &  5/26 \\ 
    Fairness neglected  & P3--P5 & 3/26 \\ 
    Context-dependent prioritization  & P22, P26 & 2/26 \\ 
     \bottomrule
    \end{tabular}
\end{table}
As depicted in Table~\ref{rq4themes}, we identified 19 final themes in answering RQ4, by collecting answers from participants about techniques they use to promote fairness, metrics to assess their fairness requirements satisfaction,  challenges that arise in validating fairness in AI software, and how fairness is traded off against performance/usability/deadlines, as well as functional feature prioritization.

\subsubsection{Techniques and Approaches for Promoting Fairness}
\imp{Participants reported a spectrum of approaches for promoting fairness in AI, ranging from data- and model-level interventions to ethical/social and organizational deliberations, with many projects lacking implemented fairness techniques.}

\paragraph{\textbf{Data-Oriented Fairness Promotion}}

\imp{Data-focused interventions were used by many participants, mainly aiming to correct imbalances and ensure representative inputs.}
Some applied data-oriented techniques such as data resampling, and balancing (e.g., P1, P5, P11, P25). 
\imp{For instance, P5 quoted the following: \\
\begin{myquote}
 {
``.. dark skin color is not common in my company...try to balance this...''.
}
\end{myquote}
}
\imp{These practices illustrate that managing fairness at the data level is often the first line of defense, directly influencing model behavior by addressing bias in the training set.}

\paragraph{\textbf{Model-Oriented Fairness Promotion}}
\imp{Some participants applied fairness-promoting techniques directly at the model level, using algorithmic or interpretability tools to guide decisions.}
\imp{For instance,} some participants (P1, P12) applied model-related fairness promotion techniques including Pareto front analysis to manage fairness–accuracy trade-offs (P1), and different explainable AI tools like LIME and SHAP to interpret model decisions, as P12 quoted: \\
\begin{myquote}
 {
``We use XAI to better understand what features intervene in the decision of models...adding the XAI results showed better accuracy. We also used LIME and SHAP to get information about attributes''.
}
\end{myquote}

\imp{This indicates that model-level interventions complement data-focused methods in a few cases, providing insights into decision pathways and enabling fine-grained fairness adjustments.}

\paragraph{\textbf{Ethical/Social and Organizational Considerations}}

\imp{Beyond technical interventions, ethical and social considerations guided participants’ fairness efforts, reflecting long-term organizational responsibility.}
For instance, P1 emphasizes ethical and social considerations, stating: ``...we have an in-depth discussion about the long-term reputational and social impacts''.

\imp{This shows that fairness promotion is not purely technical; organizational and societal factors influence how requirements are interpreted and implemented.}

\paragraph{\textbf{Lack of Implemented Fairness Techniques}}

\imp{Despite awareness of fairness issues, many participants (e.g., P4, P13, P26) reported having no formal methods or techniques to promote AI fairness, relying solely on data observation (``No techniques; just data observation'': P13) or informal practices (``No techniques or protocols'': P4).}

\imp{This highlights a significant gap in operationalized fairness practices, suggesting that awareness alone is insufficient without concrete tools or processes to enforce fairness systematically.}

\imp{Table~\ref{techniques} illustrates the different techniques/tools used by our participants to promote fairness for each of the corresponding themes and total participants reported in Table~\ref{rq4themes}}.

\begin{table}[h!]
\centering
\caption{\imp{Techniques / Approaches for Promoting Fairness Across Participants}}
\label{techniques}
\begin{tabular}{p{5cm}|p{8cm}}
\hline
\textbf{\imp{Theme (\#P)}} & \textbf{\imp{Techniques / Approaches}} \\
\hline
Data-oriented techniques (7) & Data resampling, data balancing \\
Model-oriented techniques (2) & Pareto front analysis, LIME, SHAP \\
Ethical / social and organizational considerations (1) & Long-term reputational and social impact discussions \\
No formal techniques (11) & Data observation only  \\
\hline
\end{tabular}
\end{table}
The results show that fairness practices vary widely, from no formal methods to active interventions at the data and model level, often complemented by ethical and organizational considerations. 

\subsubsection{Metrics for Validating Fairness Requirements}
Participants employed a range of approaches to validate fairness, from \imp{explicit group and individual fairness metrics} to indirect proxies, with some teams conducting limited or ad-hoc evaluations, reflecting varying levels of rigor and awareness (see the corresponding final themes in Table~\ref{rq4themes}).

\paragraph{\textbf{Explicit Group and Individual Metrics}}

\imp{Some teams applied explicit metrics at the group or individual level to directly measure fairness and identify disparities across subgroups (e.g., P1, P22), such as statistical parity, equalized odds, calibration, and individual fairness.}

These metrics are often supported by tools like Fairlearn (e.g., ``...used statistical parity...demographic parity...equalized odds... disparate impact ratio...calibration by group...individual fairness, using Fairlearn from Microsoft to visualize trade-offs between fairness and accuracy...'': P1; ``...equal opportunity of all demographic parity'': P22). 

\imp{This demonstrates that explicit, structured measurement enables teams to systematically quantify fairness, providing actionable insights and facilitating evidence-based mitigation.}

\paragraph{\textbf{Limited / Absent Fairness Evaluation}}

Limited / absent fairness evaluation captured cases (e.g., P4, P18, P20, P21) where formal metrics were not applied or only ad-hoc checks were performed (e.g., ``we didn’t have any metrics''; P20: ``No metrics... only accuracy'': P4; ``Chatbot is fair or not through A/B tests...'': P21).

\imp{This indicates that without formal evaluation frameworks, fairness practices may be inconsistent or reliant on subjective judgment, limiting accountability and reproducibility.}

\paragraph{\textbf{Standard Performance Metrics as Fairness Proxy}}

\emph{Standard performance metrics as fairness proxy} reflected various teams (e.g., P6, P8, P9, P11, P12, P18, P22) that rely on general model metrics like accuracy, F1-score, or confusion matrix to infer fairness indirectly (e.g., ``...ROC curve, confusion matrix...'': P6; ``Accuracy, precision, AUC, F1-score'': P12). 

\imp{This approach suggests that fairness is often assessed implicitly through the assessment of the overall AI model performance, which may miss subgroup disparities and fail to capture specific fairness concerns.}


\imp{Table~\ref{metrics} further illustrates the different metrics used to evaluate the AI/ML model fairness, associated with each of the corresponding themes defined in Table~\ref{rq4themes}.}

\begin{table}[h!]
\centering
\caption{\imp{Metrics Used for Evaluating Fairness Requirements Across Participants}}
\label{metrics}
\begin{tabular}{p{6cm}|p{7cm}}
\hline
\textbf{\imp{Metric Theme (\#P)}} & \textbf{\imp{Representative Metrics / Tools}} \\
\hline
Explicit group and individual metrics (2) & Statistical parity, equalized odds, calibration, individual fairness, Fairlearn \\
Limited / absent fairness evaluation (8) & No formal metrics or only ad-hoc checks \\
Standard performance metrics as fairness proxy (16) & Accuracy, F1-score, confusion matrix, ROC curve \\
\hline
\end{tabular}
\end{table}

Overall, Participants’ approaches to validating fairness ranged from using explicit group-level and individual-level metrics, to relying on standard model performance metrics as proxies, to having limited or ad-hoc evaluations, highlighting a spectrum of rigor and awareness in fairness assessment, \imp{with implications for the consistency, accountability, and completeness of bias detection across teams.}

\subsubsection{Challenges in Validating Fairness in AI Systems}
\imp{Participants reported that validating fairness in AI systems is constrained by conceptual, data, model, contextual, and evaluation-related limitations, highlighting the multifaceted complexity of operationalizing fairness.}

\paragraph{\textbf{Conceptual ambiguity in fairness definitions }}

\imp{Unclear or conflicting definitions of fairness make it difficult to establish consistent evaluation criteria across teams and projects.}
For instance, P1 shared an``...ambiguity in defining fairness'', claiming that ``fairness is often seen as a soft requirement.''. P21 added: ``...fairness took different meanings across team... no clear metrics...''. 

\imp{This shows that without a shared understanding of fairness, teams struggle to translate abstract ethical goals into actionable requirements and metrics.}

\paragraph{\textbf{Data Quality and Representation Challenges}}

\imp{Most of the reported data challenged were data-driven (e.g., P3, P5, P10--P12, P17, P22).}
For instance, P3, P5, P11, and P22 reported on outdated or imbalanced datasets (``...data is usually old...'': p3 ; ``...we don't have enough female people or black or white people...'':P5; ``data imbalance mostly'':P11; ``...ensuring enough data from all classes is a big challenge...'': P22). 

\imp{These examples underscore that fairness evaluation is only as reliable as the underlying data, making preprocessing and data augmentation critical for meaningful assessments.}

\paragraph{\textbf{Model-related challenges}}
\imp{Four out of the \total participants (P6, P8, P24, P26) reported model-related challenges faced in validating fairness in AI systems.}
For instance, P6 and P24 shared feature entanglement and model interpretability issues (``Neural network uses everything in the image...'':P6; ``...LLM behavior is the main challenge...'':P24). 

\imp{This illustrates that even with high-quality data, model complexity and opacity can prevent accurate assessment of fairness outcomes.}

\paragraph{\textbf{Contextual and organizational challenges}}

Contextual and organizational challenges (e.g., P9, P13, P15) reflected variable fairness requirements across contexts. 
\imp{For instance, P15 quoted: \\
\begin{myquote}
 {
``…different data from different contexts always impacts fairness; context-dependent…''
}
\end{myquote}
}
\imp{These challenges highlight that fairness validation is influenced not only by technical factors but also by strategic decisions and organizational norms.}

\paragraph{\textbf{Evaluation framework challenges}}

Evaluation challenges also highlighted the lack of consistent assessment systems.
\imp{For instance, P25 shared that within the organization he works for, they}
``...don’t have consistent evaluation systems''.

\imp{This demonstrates the need for robust, repeatable, and standardized evaluation tools to ensure fairness can be reliably measured and monitored.}

\imp{Overall, the challenges participants reported reveal that fairness validation is not a single-point technical problem, but a layered issue spanning unclear definitions, data limitations, model opacity, shifting contexts, and absent evaluation infrastructure. Addressing fairness meaningfully therefore requires coordinated efforts across conceptual, technical, and organizational dimensions, rather than isolated fixes at any one level.}

\subsubsection{Trade-offs Between Fairness and Project Goals (Performance, Usability, Deadlines)}

\imp{We observe three distinct ways in how teams negotiate trade-offs between fairness and project goals such as performance, usability, and deadlines: \emph{fairness as a core value}, \emph{fairness as negotiable}, and \emph{fairness as excluded from decision-making}. }

\paragraph{\textbf{Fairness as a core value}}

Some participants emphasized fairness as a core value (e.g., P11, P15, P17).
For instance, P17 said:  \\

\begin{myquote}
 {
``Fairness comes first... then performance... then extend deadline if really needed''
}
\end{myquote}
and P15 added: \\

\begin{myquote}
 {
``Fairness from data perspective always comes first... If I cannot guarantee it, I extend deadlines.''
}
\end{myquote}

\imp{These accounts illustrate a fairness-prioritization strategy, rarely adopted by AI/ML practitioners in organizations where fairness is embedded as a core value, who treat it as a non-negotiable requirement regardless of delivery pressure. In this scenario, the primary trade-off method is timeline adjustment, where deadlines are extended to allow sufficient fairness validation before deployment.}

\paragraph{\textbf{Fairness as Negotiable}}
Some participants (e.g., P8--P10, P12, P24, P26) treated fairness as negotiable or secondary, sometimes acknowledging it but not necessarily addressing it effectively, balancing it with competing demands, as in ``...performance then deadline... then think about fairness and ask costumer for deadline extension'' (quoted P26), and ``Delivery deadline but inform supervisor/customer that some fairness issues are unresolved'' (P12).

\imp{This reflects a negotiated fairness strategy adopted by some AI/ML practitioners operating under strict delivery and performance constraints, who treat fairness as secondary but not entirely dismissable. In this scenario, trade-offs are navigated through two concrete methods: extending deadlines when fairness issues are too significant to ignore, or delivering on time while formally communicating unresolved fairness limitations to supervisors or customers.}

\paragraph{\textbf{Fairness Excluded from Decision Making}}

Most of the participants described fairness as excluded (e.g., P1, P3--P5, P13), focusing instead on other goals, such as software performance (e.g., P3, P5, P10, P26), meeting deadlines (P1, P13), and software usability (P7).

For instance, P5 quoted: \\ 
\begin{myquote}
 {
``We all always take more time, so what we do is talk with the client...we don't measure fairness...our big concern is the performance''.
}
\end{myquote}
and P13 added: \\
\begin{myquote} 
 {
``Fairness is the last priority...deadline is what should be prioritized first...'' (P13).
}
\end{myquote}

\imp{These findings highlight a fairness-exclusion strategy, predominantly adopted by AI/ML practitioners working under client-driven performance and delivery constraints, who treat fairness as the lowest priority or omit it entirely from decision-making. In this scenario, no fairness-specific trade-off method is applied; instead, practitioners redirect focus toward performance optimization and deadline compliance, with fairness addressed only if explicitly required by the client.}


\subsubsection{Trade-offs Between Fairness and Functional Feature Prioritization}

\imp{Teams navigate trade-offs between fairness and functional feature delivery in ways that reflect client demands, organizational priorities, and situational relevance.}

\paragraph{\textbf{Functional Features Prioritized Over Fairness}}
\imp{Many teams prioritized functional feature delivery over fairness, treating fairness as secondary unless explicitly required.}
Many prioritized functional feature support over fairness (e.g., P1, P9, P10, P13), citing client requirements or lack of explicit fairness mandates, where P1 stated that ``unless fairness was explicitly framed as a product requirement, it would be deprioritized'', and P13 added: ``feature prioritization comes first, 100\%, fairness is a cherry on top of the cake''. 

\imp{These observations illustrate a functionality-first strategy, mostly adopted by AI practitioners working under client-driven feature requirements, who deprioritize fairness in the absence of explicit mandates. In this scenario, no fairness-specific trade-off method is applied. Instead, fairness is treated as an optional enhancement, addressed only when explicitly framed as a product requirement by the client.}

\paragraph{\textbf{Fairness Prioritized or Balanced with Functional Features}}
Some teams treat fairness as a priority or balance it equally with functional requirements (e.g., P7, P15--P17, P24).
\imp{For instance, while P7 and P15  (``make sure fairness comes first...then features after'': P7; ``still fairness comes first, then we support missing functionalities'': P15) prioritize fairness over supporting missing functionalities, P16 and P17 lean towards an equal balance between both (``missing functionality and fairness, both'': P16, P17), integrating it into the workflow.}
Further, P7 quoted the following:  \\
\begin{myquote}
 {
``make sure fairness comes first...then features after''.
}
\end{myquote} P15 added: \\

\begin{myquote}
 {
``still fairness comes first, then we support missing functionalities''.
}
\end{myquote}

\imp{These examples highlight a fairness-inclusive strategy, adopted by some AI/ML practitioners who treat fairness as either a prerequisite or an equal counterpart to functional delivery. In this scenario, two concrete methods are used: sequencing fairness validation before addressing missing functionalities, or integrating fairness and functional requirements in parallel as equally weighted priorities throughout the workflow.}


\paragraph{\textbf{Fairness Neglected}}
\imp{In some cases, fairness is overlooked entirely, with functional priorities dominating AI decisions (e.g., P3--P5).}
\imp{For instance, P4 said:} \\
    \begin{myquote}
     {
    ``.... we didn't have a requirement for fairness, so there wasn't a trade off.''.
    }
    \end{myquote} 
    
\imp{These accounts illustrate a fairness-absent strategy, adopted by a few AI/ML practitioners operating in contexts where fairness is never raised as a requirement, leaving functional priorities to dominate AI development decisions entirely. In this scenario, no trade-off method is applied, as fairness is not recognized as a competing concern in the first place.}

\paragraph{\textbf{Context-dependent prioritization}}
Some decisions were context-dependent, addressing fairness only when required or relevant to a functional feature (e.g., P22, P26).
For instance, P22 quoted: 
\begin{myquote}
     {
    ``If extra feature has relation with the fairness, then I prioritize fairness. Otherwise, I work on the missing functionality then reflect on the fairness''
    }
\end{myquote}

\imp{These accounts illustrate a context-driven strategy, adopted by two AI researchers who address fairness conditionally, based on its relevance to the functional feature at hand. In this scenario, the trade-off method consists of evaluating each feature individually, prioritizing fairness when a direct link to it is identified, and deferring it otherwise until functional delivery is complete.}

Overall, fairness is often considered secondary to delivering functional software features, where the latter is prioritized to satisfy client needs, and the former is frequently overlooked in practice unless explicitly required or contextually relevant.
\imp{These patterns reveal a spectrum of trade-off strategies, showing that the way fairness is handled depends on explicit requirements, organizational values and situational relevance, painting a coherent picture of how teams navigate competing priorities in real projects.}

\begin{summarybox}[RQ4 Summary]
Participants showed wide variation in fairness practices, from no formal methods to data/model-level techniques and the application of proper fairness metrics. Our results show that fairness validation is inconsistent, often relying on proxies or ad-hoc checks, with major challenges in adopting fairness definitions and the data/model use. Further, fairness was frequently deprioritized against the AI/ML model's effectiveness, strict delivery deadlines, and software functional feature support, underscoring the need for clearer policies and structured, consistent evaluation.
\end{summarybox}

\paragraph{\textbf{Quantitative Summary of RQ4 Observations Across Participants}}

\imp{As shown in Table~\ref{rq4themes}, data-oriented fairness promotion was reported by seven participants, including a senior data science manager from North America, a highly senior professor from Africa/North America, and mid-level AI engineers and data scientists from South America. 
Model-oriented fairness promotion was noted by two participants, an early-career computer vision engineer from Africa and a senior data science manager from North America. 
Ethical, social, and organizational considerations were raised by a senior data science manager from North America. 
Lack of implemented fairness techniques was reported by eleven participants, spanning early- to highly senior-level AI/ML researchers, engineers, and R\&D specialists across Asia, Africa, and North America.}

\imp{Explicit group and individual fairness metrics were reported by three participants, including a senior data science manager from North America and a senior AI researcher from Asia.
Limited or absent fairness evaluation was reported by 12/26 participants, including early- to mid-level A/ML engineers and developers, and R\&D specialists across North America, Europe, Asia, and Africa. 
Finally, standard performance metrics used as a proxy for fairness were reported by 11/26 participants, including senior AI researchers /data scientists, and early- to mid-level AI engineers, computer vision engineers, and research fellows across Asia, Europe, and Africa.}

\imp{
Data quality and representation challenges were frequently reported by 10/26 participants, including early- to senior-level AI engineers, AI/ML researchers, and computer vision engineers across Asia, North America, South America, and Africa. Model-related challenges were reported by four participants, including early-career to senior ML software developers, AI engineers, and R\&D specialists across South America and Asia. 
Contextual and organizational challenges were reported by four participants, including early- to mid-level AI researchers, engineers, and consultants across Asia, Europe, North America, and Africa. 
Evaluation framework challenges were reported by three participants, including mid-level data scientists from South America and Asia.
Finally, conceptual ambiguity in fairness definitions was reported by two participants, representing a senior data science manager from North America and an early-career R\&D researcher from Asia.}

\imp{
The exclusion of  fairness from decision making was reported by the majority of our participants (16/26), including early- and senior-level AI/ML researchers and engineers, ML software developers, computer vision engineers, and R\&D specialists across Asia, North America, South America, and Europe.
Fairness as negotiable was reported by six participants, including early to mid-level AI/ML researchers/ engineers/ software developers, across Asia and Africa, while fairness consideration as a core value was highlighted by four participants, including mid-level AI consultants and ML researchers across Asia and North America. 
}

\imp{Functional features prioritization over fairness was frequently reported by 16/26 participants, including early-to senior-level AI researchers and computer vision engineers, across North America, Asia, Europe, and Africa.
Fairness prioritized over functional feature delivery or balanced prioritization was less commonly adopted by five early to senior AI/ML researchers / consultants and software engineers across Asia, Europe, and North America. 
Fairness being completely neglected was less commonly reported by three participants, including a mid-level AI engineer and ML software developer, across North and South America, along with context-dependent prioritization, reported by two participants, including mid- to senior-level AI researchers and engineers located in Asia and South America.}

\begin{summarybox2}[Quantitative Summary of RQ4]
\imp{Quantitative analysis of RQ4 findings shows that approaches to promoting and validating fairness, as well as fairness-related challenges and trade-offs, are widely distributed across seniority levels, from early- to highly senior participants. Patterns span diverse roles, including AI/ML researchers, engineers, data scientists, R\&D specialists, consultants, software developers, and project managers. Geographically, these patterns cover multiple continents, including North America, South America, Europe, Asia, and Africa. Overall, both fairness promotion techniques (data- and model-oriented), evaluation practices (explicit metrics, limited assessment, proxies), challenges (data, model, contextual, and conceptual), and prioritization decisions (functional vs. fairness-focused) are broadly shared across roles, experience levels, and regions, indicating no restriction to specific positions, seniority levels, or locations, while more structured or ethically driven approaches remain less common.}
\end{summarybox2}
\section{Discussion}\label{disc}
In this section, we discuss the insights from our research, \imp{along with the corresponding practical implications and our recommendations to practitioners.}

\subsection{Insights And Practical Implications}

\imp{Our findings show that} 
fairness definition (RQ1 in ~\ref{rq1}) in AI/ML systems is nuanced and inconsistent among AI/ML practitioners, along with the corresponding evaluation metrics \imp{(RQ4 in ~\ref{rq4})}, based on collected answers from our interviews \imp{across diverse roles, expertise levels, and geographical sites of the projects,} and prior empirical studies~\cite{de2025software,deng2023investigating,ferrara2024fairnessinterview,holstein2019improving,pant2025navigating,rakova2021responsible,ryan2023integrating,ryanfairness,shin2019role,smith2025pragmatic}. 

\imp{While several early- to senior-level interview participants defined fairness as absence of favoritism or discrimination in the corresponding model decisions (see Section ~\ref{rq1}), aligning with prior studies~\cite{de2025software,ferrara2024fairnessinterview, pant2025navigating,ryan2023integrating,shin2019role}, the majority defined it in terms of data quality (further associating fairness issues with poor data quality such as imbalanced datasets, inadequate representation of minority groups, and non-careful handling of sensitive features, aligning with prior findings~\cite{de2025software,deng2023investigating,holstein2019improving,pant2025navigating,ryan2023integrating}), mainly among early to mid career AI/ML practitioners, while only a few mid- to senior-level practitioners defined it in terms of ethical considerations, we noticed unfamiliarity with the concept of fairness. We also although practitioners demonstrated understanding as the interview progressed (e.g., ``first time I hear about fairness'': said P20).
This inconsistency in defining AI fairness reflects a lack of shared understanding among practitioners, leading to inconsistency in fairness interpretation and the corresponding practices.
Further, one participant (P2, a mid-level data scientist) shared the adoption of two fairness definitions from model and ethics perspective, reflecting overlapping understandings of fairness across different perspectives, which can create contradictions and undermine the overall project goals~\cite{baresi2023understanding,demirchyan2025algorithmic,ramadan2025towards,ferrara2024fairness,ryan2023integrating,ryanfairness,chen2024fairness}. 
Overall, achieving agreement on which fairness definitions to adopt is essential before advancing in the AI development lifecycle, as attempting to satisfy multiple definitions simultaneously can create contradictions and undermine the overall project goals~\cite{baresi2023understanding,demirchyan2025algorithmic,ramadan2025towards,ferrara2024fairness,ryan2023integrating,ryanfairness,chen2024fairness}. 
}

\imp{Additionally, fairness requirement documentation is largely inconsistent or absent (see RQ2 in \ref{rq2}), where the two largest groups of observed themes either indicate the inconsistent use of shared digital platforms (mainly early- to mid-level AI/ML practitioners) or a lack of formal documentation altogether, ranging from early- to senior-level AI/ML practitioners. The inconsistency and/or the absence of fairness documentation reflects a lack of standardized practices, resulting in gaps in potential traceability, accountability, and shared understanding across teams.}

\imp{Further, in interpreting fairness requirements, early- to senior-level AI/ML practitioners most shared the involvement of technical and domain-expert stakeholders, followed by cross-functional and operational teams (which aligns with the recommended stakeholder involvement strategy in the literature~\cite{nguyen2025gray,singh2022fair,lu2022towards,ferrara2024fairness,mccormack2024ethical,pant2025navigating,pham2025fairness,ryan2023integrating,voria2024catalog}), with little involvement of management and leadership stakeholders reported by early- to mid-level AI/ML practitioners. 
This indicates that fairness decisions rely heavily on technical contributors, which may limit organizational oversight and broader governance.}

\imp{Fairness promotion techniques were unevenly applied across projects, where, the majority of participants from all seniority levels reported lack of implemented fairness techniques (also aligning with prior findings~\cite{de2025software}), while others mostly rely on data-oriented approaches, representing early- to mid-level AI/ML practitioners. 
This indicates a limitation of fairness-related efforts in promoting AI software, leaving many projects without formal techniques to ensure fairness, which may compromise the reliability and accountability of AI/ML systems.}

\imp{Fairness validation metrics vary widely across projects. Explicit group and individual fairness metrics were reported by a few early- to mid-level AI participants. The rest of participants was near evenly split between reporting limited or absent fairness metrics by early- to mid-level AI/ML practitioners, and using standard performance metrics as fairness proxy (aligning with prior findings~\cite{de2025software}) by early- to senior-level participants. This indicates that fairness is often not explicitly measured, with many practitioners relying on general performance metrics or no metrics at all, which may lead to unrecognized biases and unreliable fairness assessments.}

\imp{Trade-offs between fairness and project goals reveal that, while some early- to mid-level AI/ML participants prioritize or balance fairness with other features in AI software development, fairness is frequently deprioritized or neglected in practice (RQ4 in ~\ref{rq4}). More in detail, when traded-off with performance or meeting deadlines, fairness is commonly excluded from decision making by AI/ML practitioners across all seniority levels, with the rest split between considering fairness as negotiable or prioritizing fairness. 
When traded-off with the support of a missing functional requirement, priority is given to the latter by most practitioners across all seniority levels, while the remaining participants involved either ignoring fairness, balancing it with other objectives, or prioritizing it when it is a functional requirement, such as a client need.
This indicates that fairness is often secondary to core project objectives, aligning with findings from prior studies~\cite{ferrara2024fairnessinterview,smith2025pragmatic,de2025software,pant2025navigating,rakova2021responsible,ryan2023integrating}. This may lead to overlooking fairness or addressing it inconsistently in AI/ML systems.
}

\minor{Recall that for each theme across all four RQs, we performed cross-cut comparisons by tracing supporting participants back to their demographic attributes in Table~\ref{demographics} (see Section~\ref{eval}). No meaningful differences emerged across roles, seniority levels, or regions, and no theme was found to be concentrated within a particular role, career stage, or geographic context. This null result is itself informative, suggesting that the immaturity of fairness practices is structural and industry-wide rather than tied to individual practitioner profiles.}

\imp{Together, these implications suggest that organizations should adopt structured and context-aware approaches to define, document, manage, apply, and validate fairness requirements in the AI/ML software, while embedding them throughout the AI lifecycle, making sure to engage stakeholders in the whole process.
More importantly, organizations should explicitly balance fairness requirements against other technical, operational, and organizational constraints.}

\subsection{Recommendations}

In light of the above insights and \imp{the corresponding implications}, AI/ML fairness should be considered throughout the entire SDLC process, from requirement elicitation to deployment and maintenance. 
Incorporating fairness early in the SDLC allows organizations to proactively address the biased AI decisions before it becomes costly in later stages.
\imp{In the following, we organize our recommendations along three dimensions: training, guidelines, and operational practices, cover the knowledge, processes, and technical steps needed to implement fairness across the SDLC. 
More in detail, \textbf{training} addresses the inconsistency / ambiguity in how practitioners define fairness and understand fairness concerns (RQ1).
\textbf{Practical guidelines} are needed to help practitioners
manage fairness in all phases of the AI lifecycle~\cite{ferrara2024fairnessinterview,de2025software} (RQ2). \textbf{Operational practices} enable the application and validation of fairness, addressing the corresponding faced challenges, along with trade-offs with other project priorities (RQ3 and RQ4). Together, these dimensions cover the knowledge, processes, and technical steps needed to implement fairness effectively across the SDLC.}

\paragraph{\imp{\textbf{Training}}}
First, training is mandatory across organizations to clarify the distinction between AI/ML model fairness and their accuracy. 
This is needed to increase the software practitioners' awareness of AI fairness and therefore encourage them to prioritize it in the AI software development lifecycle.
\imp{More specifically, training should include i) educating practitioners about AI/ML fairness and what it stands for, highlighting common misunderstandings such as confusing fairness with model accuracy, and ii) sharing practical examples of fairness concerns observed in real projects. It should be delivered through in-company workshops or hands-on sessions by AI/ML fairness experts, given to all new staff and repeated periodically for existing team members, with completion tracked by team leads or HR.}

\paragraph{\textbf{\imp{Practical Guidelines}}}

\imp{First, practitioners need to proactively manage trade-offs between fairness and other project objectives, including accuracy, system functionality, user experience, and organizational priorities~\cite{holstein2019improving,rakova2021responsible,pushkarna2022data,baresi2023understanding,demirchyan2025algorithmic,ferrara2024fairness}, by incorporating early planning (e.g., scheduling trade-off assessments during project scoping) and setting  clear policies such as specifying which objectives take priority when fairness conflicts with other objectives. 
}

\imp{Integrating fairness across the AI pipeline requires structured, phase-specific practices.
To do so, organizations should first define fairness, as this definition will serve to translate fairness concerns into concrete, measurable requirements that will afterwards be formulated into validation metrics to assess and validate AI model fairness.}

\imp{Additionally, fairness requirements need to be properly documented, with documentation maintained and potentially updated as new fairness requirements are added or existing ones are refined.}

\imp{Given the inconsistency in stakeholder involvement in AI fairness, practitioners should adopt an early and broad stakeholder involvement in the SDLC, with AI/ML practitioners leading the development of these guidelines, consulting domain experts, managers, policymakers, and legal advisors. This ensures that fairness requirements are effectively defined, documented, refined (when needed), interpreted, applied, and finally validated in AI software, while remaining aligned with practical, legal, and business considerations.}

\imp{Further, organizations should consider incorporating human supervision when developing AI/ML systems, ensuring that model behavior and outputs are actively monitored for fairness throughout the AI lifecycle. This aligns with the Fair CRISP-DM framework~\cite{singh2022fair}, which embeds human oversight into both data preparation and model training stages to mitigate biases early rather than relying solely on post-training fixes.
}

\paragraph{\textbf{\imp{Operational Practices}}}
We suggest using systematic bias detection, monitoring, and mitigation to guarantee clarity, traceability, and consistent application of fairness across the SDLC.
\imp{For instance, AI/ML practitioners should embed continuous fairness checks at i) the data pre-processing step to detect (and eventually mitigate) training data bias~\cite{singh2022fair} such as underrepresentation of certain groups or skewed label distributions, before they propagate through the model, and ii) the model training step, applying, for instance, fairness constraints or regularization terms directly to the loss function, penalizing the model when it learns patterns that produce disparate treatment across demographic groups.
Additionally, we strongly suggest to support audit trails (also suggested in some prior research~\cite{singh2022fair,ferrara2024fairnessinterview}) to ensure traceability and accountability, including the record of data-related activities such as data pre-processing, balancing and augmentation, along with monitoring model training and inspecting outputs during post-processing~\cite{holstein2019improving,pham2025fairness,ferrara2024fairness,ferrara2024refair}}.

\subsection{\imp{Key Challenges Framework}}

\imp{Table~\ref{model} synthesizes our findings into a framework of interconnected challenges (with corresponding descriptions, related research questions, root causes, and their impact on fairness outcomes) across multiple dimensions. 
As shown in the table, fairness challenges in AI systems arise from both cognitive and technical gaps. Challenges begin with \emph{Conceptual Ambiguity and Cognitive Gaps}, where a lack of shared understanding, driven by several causes such as skill gaps and fairness deprioritization, results in inconsistent and superficial fairness practices. This weak conceptual grounding leads to both \emph{Weak Formalization and Documentation}, where fairness concerns are not translated into explicit and measurable requirements, and \emph{Evaluation and Metric Challenges}, where unclear standards and lack of expertise hinder meaningful fairness assessment. The inconsistency in defining fairness, mainly due to lack of expertise or training gaps, leads to operationalizing it narrowly, primarily regarding fairness from either a data perspective (\emph{data-centric fairness challenges}) or a model perspective (\emph{model-level fairness challenges}). The latter restricts fairness assessment primarily to model outcomes due to several reasons such as overreliance on algorithms and insufficient interpretability (i.e., practitioners do not understand the AI model behavior), potentially leading to bias propagation, inconsistent model predictions, and potentially other overlooked issues. The former, however, restricts fairness to data due to related issues (i.e., data quality gaps) and weak governance, potentially leading to treating fairness as a data preprocessing task.
\noindent These challenges are further amplified by \emph{Fragmented Stakeholder Involvement}, caused by a lack of governance structures (e.g., no clear rules or assigned roles to guide AI fairness), leading to inconsistent implementation and interpretation of fairness. All this contributes to \emph{Process and Lifecycle Integration Gaps}, where fairness is not systematically embedded in the AI lifecycle, often resulting in reactive practices rather than proactive ones (e.g., fixing bias only after deployment rather than designing fairness checks earlier in the SDLC). Among the reasons this occurs are \emph{Trade-off Dilemmas and Practical Constraints}, where fairness is often deprioritized when traded off against other objectives such as functional requirements and delivery deadlines, potentially leading to fairness being completely ignored in AI system development.}

\begin{longtable}{p{2.4cm} p{4.5cm} p{3cm} p{2.7cm}}
\caption{\imp{A Framework of Key Challenges in Operationalizing AI Fairness}} \\
\hline
\textbf{Challenge} & \textbf{Description [RQ]} & \textbf{Root Causes} & \textbf{Outcomes} \\
\hline
\endfirsthead

\multicolumn{4}{c}%
{{\bfseries \tablename\ \thetable{} -- continued from previous page}} \\
\hline
\textbf{Challenge Dimension} & \textbf{Description [RQ]}  & \textbf{Root Causes} & \textbf{Outcomes} \\
\hline
\endhead

\hline \multicolumn{4}{r}{{Continued on next page}} \\
\endfoot

\hline
\endlastfoot

\textbf{Conceptual \text{Ambiguity \&} \text{Cognitive Gaps}} &
Lack of shared understanding of fairness, manifesting in inconsistent definition of fairness \textbf{[RQ1]}.
&
Skill gaps, training gaps, low to no fairness prioritization
& \text{Inconsistent and} \text{superficial fairness} practices \\
\hline
\textbf{Data-Centric Fairness \text{Challenges}} &
Fairness primarily regarded from data quality perspective \textbf{[RQ1--RQ4]}, with limited attention to broader system-level issues &
data quality gaps, weak governance, training gaps  &
\text{Partial mitigation,} fairness mainly dealt with as a data preprocessing task \\
\hline
\textbf{Model-Level Fairness \text{Challenges}} &
Fairness primarily interpreted from model outcome perspective \textbf{[RQ1, RQ3, RQ4]} &
overreliance on algorithms, insufficient interpretability, training gaps, lack of expertise  &
Bias propagation, overlooked data \text{issues, inconsistent} \text{model predictions} \\ \hline
\textbf{Weak \text{Formalization} \& Documentation} &
Limited translation of fairness concerns into explicit, measurable, and documented requirements \textbf{[RQ2]} &
Unclear / lack of standards, low quantification, limited / absent metrics, low prioritization &
Ad hoc practices; lack of traceability and accountability \\
\hline
\textbf{\text{Process \&} \text{Lifecycle} \text{Integration Gaps}} &
Fairness is not systematically embedded across the SDLC and is often handled in isolated or late stages \textbf{[RQ2--RQ4]} &
Weak formalization, low to no documentation, fragmented \text{stakeholders, low} prioritization time constraints &
Reactive rather than proactive fairness practices \\
\hline
\textbf{\text{Evaluation /} Metric \text{Challenges}} &
Limited use of explicit fairness metrics, with a reliance on performance proxies or absence of fairness evaluation \textbf{[RQ4]} &
Tooling gaps, lack of expertise, unclear standards &
unvalidated fairness, untrustworthy results, skewed insights \\
\hline
\textbf{Fragmented Stakeholder Involvement} &
\text{Unclear, weakly-defined,} \text{limited, or inconsistent} \text{stakeholder involvement in} promoting fairness \textbf{[RQ2]} &
Lack of governance structures &
Inconsistent implementation and interpretation of fairness \\
\hline
\textbf{Trade-off \text{Dilemmas \&} Practical \text{Constraints}} &
Unresolved trade-offs between \text{fairness and performance,} \text{usability, or timelines \textbf{[RQ4]}} &
\text{Lack of decision} frameworks, delivery \text{pressure, performance} pressure
 & Fairness compromised or ignored \\
 
\label{model}
\end{longtable}
\subsection{Threats to Validity}
The use of semi-structured interviews to assess practitioners’ awareness and implementation of fairness in AI/ML models may introduce several internal, external, and construct threats to validity, as follows:

\subsubsection{Internal Threats}
Coding and interpreting the collected transcripts from the \total interview participants may have resulted in inconsistency in reported results and bias in consolidating answers as interviews were conducted by two researchers from different backgrounds. 
This potential bias may unintentionally influence participants' understanding of fairness requirements in AI/ML software, shifting it toward the researchers' own ethical expectations or assumptions, possibly leading to inconsistencies in reporting codes. To mitigate this, both researchers cross-reviewed each other's codes independently.
\imp{Additionally, the absence of post-hoc member checking may affect the credibility of our findings. As none of the participants agreed to be re-contacted to verify transcripts or thematic interpretations, we relied on real-time clarification during interviews as a way to partially mitigate this threat to validity.}

\subsubsection{External Threats}
Although our qualitative analysis provides valuable insights into practitioners' awareness and practices of fairness testing in AI models, our findings are not intended for statistical generalization due to the limited sample size. To mitigate this, we interviewed software practitioners from different organizations (universities, research labs, and companies), application domains (e.g., e-commerce, healthcare, banking, and computer vision), countries, different genders, and different levels of expertise, ensuring diversity in backgrounds and perspectives. 
\minor{Further, as our sample reflects predominantly small-to-medium-sized team contexts (4--12 members,  see Section~\ref{targetpopul}), our findings may not fully generalize to large-enterprise settings where fairness governance structures (e.g., dedicated responsible AI teams or formal ethics review boards) may differ substantially. This reflects a structural sampling boundary of our study, partially addressed by the diversity of our sample in terms of roles, seniority levels, and regions, broadening the range of fairness-related perspectives.}
Our qualitative analysis follows established qualitative sampling guidelines~\cite{baltes2022sampling}, which enhance the relevance of our insights. Further, the lack of systematic data on company size and organizational maturity, due to incomplete information provided by participants, limits the generalizability of our findings. We mitigated this limitation by reporting team sizes where available and explicitly noting the resulting constraints on generalizability.
\imp{Additionally, recruitment via LinkedIn and personal contacts may introduce coverage and self-selection bias, as volunteers may be more interested in fairness than the broader practitioner population. To mitigate this, we ensured diversity across organizations, domains, countries, genders, and levels of expertise, and applied inclusion criteria requiring hands-on AI/ML experience.}
\imp{Additionally, because fairness is a socially desirable topic, participants' responses may be influenced by social-desirability or impression-management effects, potentially leading them to overstate their awareness or application of fairness practices. To mitigate this, we informed our participants that their responses will remain anonymized, to encourage honest answers based on their actual experience.}

\subsubsection{Construct Threats}
Our participants come from diverse countries and professional backgrounds, with varying levels of English proficiency. Questions were rephrased to accommodate different English levels, and for two participants, the interviews were conducted in French.
This can lead to variations in interpreting key concepts and semantics shift. To mitigate this threat, we carefully rephrased questions for participants with limited English proficiency and also conducted interviews in fluent French, preserving the exact semantics of the original English questions. Additionally, although we did not conduct triangulation to validate our findings through multiple data sources, we mitigated this by comparing our results with prior empirical studies, highlighting both common patterns and new insights emerging from our work. 

\subsection{Limitations}
Despite following a thematic analysis~\cite{clarke2017thematic} and empirical standards~\cite{baltes2022sampling} with shared representative quotes from our participants, along with the corresponding initial codes, and the candidate, refined, and final themes, some nuances may have been missed. 

\minor{Additionally, our participants sample reflects predominantly small-to-medium-sized team contexts (4--12 members, see Section~\ref{targetpopul}), as systematic data on company size and organizational maturity was not available for all participants. Large-enterprise environments where fairness governance may involve dedicated responsible AI teams, formal ethics review boards, or regulatory compliance obligations are not represented in our sample and may exhibit different practices.}

Fully understanding fairness considerations in AI systems remains challenging due to their complexity and varied organizational practices. Although interview questions were carefully designed to properly address each of our research questions RQ1--RQ4 (see Section~\ref{rqs}), our study does not include a formal validation method to assess the extent to which our interview questions cover the addressed research questions.

\section{Conclusion \& Future Directions}\label{conc}
Our study on fairness in AI/ML projects through \total interviews with software practitioners across various contexts and different experiences spanning different countries reveals that while awareness of AI/ML fairness is generally present, its understanding and application in real-world problems remain uneven and inconsistent across projects. Similarly, AI/ML experts reflect a noticeable knowledge gap in defining fairness and related requirements, and show inconsistency in documenting, integrating in the SDLC, validating, and evaluating them, which can undermine the effectiveness of fairness interventions.
Further, participants always face trade-offs between fairness and other objectives, such as supporting missing functionalities, meeting strict delivery deadlines, and AI model effectiveness, often leading to fairness deprioritization. 
\imp{To further support these insights, we modeled the key challenges in a framework that systematically organizes the diverse findings observed in practice, highlighting the main obstacles practitioners face when assessing AI fairness.
Our findings also indicate that these patterns are consistent across practitioners of different roles, levels of expertise, and locations of the different AI projects they worked on, suggesting that the challenges are broadly shared rather than restricted to specific contexts.}
\minor{While these patterns appear consistent across the diverse contexts represented in our sample, exploring how they manifest in large-enterprise settings, where additional governance structures and regulatory pressures may shape fairness practices differently, remains a promising direction for future work.
}
Overall, our findings highlight the need for standardized definitions, metrics, and formal processes to systematically guide the integration of fairness throughout AI/ML projects, supporting responsible, accountable, and ethically aligned systems.

\clearpage
\appendix
\section{\imp{Interview Guide}}\label{appendix}


\subsection{Introduction and Background}

This section introduces the purpose of the interview, followed by questions about the participants’ professional context. The Interviewer aims to gather details about the Interviewee’s experience with fairness requirements in AI. More in detail, the purpose of this study is to investigate how professionals identify and validate the fairness requirements.
Questions on the Education and Experience background of the Interview Participants:
\begin{itemize}
    \item How many years of industry experience do you have working with AI or ML  systems?
    \item What position do you currently hold in your team?
    \item What is your highest level of education completed?
\end{itemize}

\subsection{Research Questions}

\begin{itemize}
    \item RQ1: How aware are AI/ML professionals of key concepts of fairness in AI/ML software
    \item development, and how fairness has been reflected in their work?
    \item RQ2: How are fairness concerns translated into fairness requirements? And how are the latter refined, documented, and interpreted by stakeholders in AI/ML software development processes?
    \item RQ3:How do fairness concerns arise and manifest early in AI/ML projects within the SDLC? And what challenges do teams face in handling them?
    \item RQ4: How do teams implement, validate, and evaluate fairness requirements during development? And how do they manage trade-offs with other project goals?
\end{itemize}

\subsection{Interview Questions}

\subsubsection{RQ1  Interview Questions:}
\begin{itemize}
    \item Q1: When you hear the term fairness in the context of software or AI development, what does it mean to you?
    \item Q2: Can you share an example from your work where software fairness was a concern or topic of discussion during the software development process? 
\end{itemize}

\subsubsection{RQ2 Interview Questions:} 
\begin{itemize}
    \item Q1: How are fairness concerns turned into requirements?
    \item Q2: Can you describe how your team approached the collection of fairness-related requirements?
    \item Q3: Can you walk us through the process where your team collected, analyzed or refined a fairness requirement during development?
    \item Q4: How are fairness requirements documented within your company / team?
    \item Q5: How were different stakeholders (e.g., developers, designers, testers, users) involved in interpreting fairness requirements in a specific context?
\end{itemize}

\subsubsection{RQ3 Interview Questions:}
\begin{itemize}
    \item Q1: Can you share a situation where fairness concerns were raised during the early stages of the project?
    \item Q2: What limitations or challenges does your team face when identifying or applying fairness in the project?
\end{itemize}

\subsubsection{RQ4 Interview Questions:}
\begin{itemize}
    \item Q1: What techniques, approaches, or procedures are used to promote fairness during software development in the project you work on?
    \item Q2:What are the metrics that your team used to validate the fairness requirements?
    \item Q3: What challenges were you facing when trying to validate the fairness requirements in AI systems?
    \item Q4: How does your team manage trade-offs between fairness and other project goals such as performance, usability, and delivery deadlines?
    \item Q5: How does your team manage trade-offs between fairness and functional feature prioritization?
\end{itemize}

\subsubsection{Reflections \& Improvements}
\begin{itemize}
    \item Q1: Is there any additional information you would like to mention that has not been covered during the interview?
\end{itemize}


\bibliographystyle{plainnat}
\bibliography{bibliography}
\end{document}